\documentclass[journal,letterpaper]{IEEEtran}
\usepackage{geometry}
\usepackage{color}
\usepackage{float}
\usepackage{amsthm}
\usepackage{amsmath}
\usepackage{amssymb}
\usepackage{graphicx}
\usepackage{url}
\usepackage{cite}
\usepackage{epsfig}
\usepackage{anysize}
\usepackage{algorithm}
\usepackage{algorithmic}
\usepackage{comment}

\DeclareMathOperator*{\minimize}{\mathrm{minimize}}
\DeclareMathOperator*{\maximize}{\mathrm{maximize}}
\DeclareMathOperator*{\st}{\mathrm{subject\;to}}
\DeclareMathOperator*{\var}{\mathrm{variables}}
\DeclareMathOperator*{\argmin}{\arg\min}

\makeatletter

\providecommand{\tabularnewline}{\\}
\floatstyle{ruled}
\newfloat{algorithm}{tbp}{loa}
\providecommand{\algorithmname}{Algorithm}
\floatname{algorithm}{\protect\algorithmname}

\theoremstyle{plain}
\newtheorem{thm}{\protect\theoremname}
\theoremstyle{plain}
\newtheorem{prop}{\protect\propositionname}
\theoremstyle{plain}

\theoremstyle{plain}
\newtheorem{lem}{\protect\lemmaname}
\theoremstyle{remark}
\newtheorem{rem}[]{\protect\remarkname}
\theoremstyle{note}

\marginsize{.8in}{.8in}{.7in}{.7in}
\newenvironment{revise}{}{}
\providecommand{\notename}{Note}
\providecommand{\remarkname}{Remark}
\providecommand{\lemmaname}{Lemma}
\providecommand{\corollaryname}{Corollary}
\providecommand{\propositionname}{Proposition}
\providecommand{\theoremname}{Theorem}

\ifCLASSOPTIONcompsoc
\usepackage[caption=false,font=normalsize,labelfont=sf,textfont=sf,labelformat=simple]{subfig}
\else
\usepackage[caption=false,font=footnotesize,labelformat=simple]{subfig}
\fi

\makeatother

\begin{document}

\title{Cache-Enabled\! Physical\! Layer\! Security\! for\! Video Streaming\! in\! Backhaul-Limited\! Cellular\! Networks\!
\thanks{The work of D. W. K. Ng was supported under Australian Research Council’s Discovery Early Career Researcher Award funding scheme (DE170100137). The work of R. Schober was supported by the Alexander von Humboldt Professorship Program. The work of V.W.S. Wong was supported by the Natural Sciences and Engineering Research Council of Canada. This work was presented in part at the \emph{IEEE Global Communications Conference (Globecom)} 2016,  Washington, DC, USA, Dec. 2016 \cite{Xiang16:CoMP}. 
}
}

\author{{Lin~Xiang,~\IEEEmembership{Student~Member,~IEEE,} Derrick~Wing~Kwan~Ng,~\IEEEmembership{Senior~Member,~IEEE,} \\ Robert~Schober,~\IEEEmembership{Fellow,~IEEE,} and Vincent~W.S.~Wong,~\IEEEmembership{Fellow,~IEEE}}

%\author{{Lin Xiang, Derrick Wing Kwan Ng, Robert Schober, and Vincent W.S. Wong}
%
\thanks{L.~Xiang and R.~Schober are with the Institute for Digital Communications, Friedrich-Alexander University of Erlangen-Nuremberg, Erlangen 91058, Germany (Email: \{lin.xiang, robert.schober\}@fau.de).}  
\thanks{D. W. K.~Ng is with the School of Electrical Engineering and Telecommunications, University of New South Wales, Sydney, NSW 2052, Australia (Email: w.k.ng@unsw.edu.au).}
\thanks{V.W.S.~Wong is with the Department of Electrical and Computer Engineering, University of British Columbia, Vancouver, BC V6T 1Z4, Canada (Email: vincentw@ece.ubc.ca).}
\vspace{-.2cm}
}

\maketitle
\pagenumbering{gobble} 

\vspace{-.2cm}
\begin{abstract}
In this paper, we propose a novel wireless caching scheme to enhance the physical layer security of video streaming in cellular networks with limited backhaul capacity. By proactively sharing video data across a subset of base stations (BSs) through both caching and backhaul loading, secure cooperative joint transmission of several BSs can be dynamically enabled in accordance with the cache status, the channel conditions, and the backhaul capacity. Assuming imperfect channel state information (CSI) at the transmitters, we formulate a two-stage non-convex mixed-integer robust optimization problem for minimizing the total transmit power while providing quality of service (QoS) and guaranteeing communication secrecy during video delivery, where the caching and the cooperative transmission policy are optimized in an offline video caching stage and an online video delivery stage, respectively. Although the formulated optimization problem turns out to be NP-hard, low-complexity polynomial-time algorithms, whose solutions are globally optimal under certain conditions, are proposed for cache training and video delivery control. Caching is shown to be beneficial as it reduces the data sharing overhead imposed on the capacity-constrained backhaul links, introduces additional secure degrees of freedom, and enables a power-efficient communication system design. Simulation results confirm that the proposed caching scheme achieves simultaneously a low secrecy outage probability and a high power efficiency. Furthermore, due to the proposed robust optimization, the performance loss caused by imperfect CSI knowledge can be significantly reduced when the cache capacity becomes large. 

\end{abstract}

\vspace{-0.2cm}

\section{Introduction}

\IEEEPARstart{T}{he} rapidly growing video-on-demand (VoD) streaming traffic in cellular networks has introduced significant challenges for  service providers as both the radio resources in the radio access network (RAN) and the capacity of the backhaul links are limited \cite{WSJ13:Verizon,Rost14MCOM:CloudRAN}. Wireless caching has been proposed to meet the stringent VoD streaming requirements in the fifth generation (5G) cellular networks \cite{Paschos16WC,Liu13TSP:CoMP,Liu14TSP:CoMP,chen16cooperative,Tao16TWC:Multicast,Xiang17TVT:CLCaching}. Different from traditional wireless networking paradigms, wireless caching is a content-centric solution for VoD streaming and intelligently exploits the fact that the content requested by users is highly correlated \cite{Breslau99Zipf}. By pre-storing the most popular files at base stations (BSs) and access points (APs), wireless caching enables quick access to these files via wireless networks and, consequently, improves the quality of service (QoS) in video streaming. Meanwhile, by reusing the cached content for transmission to multiple users, the backhaul traffic is significantly reduced \cite{Paschos16WC}. Recently, caching has also been exploited as a physical layer mechanism to facilitate cooperative multiple-input multiple-output (MIMO) transmission \cite{Liu13TSP:CoMP,Liu14TSP:CoMP,chen16cooperative,Tao16TWC:Multicast} and cross-layer resource allocation \cite{Xiang17TVT:CLCaching}. These schemes effectively exploit the multiplexing and diversity gains introduced by caching for spectral efficiency enhancement in the RAN as well as energy savings in the entire cellular network. Therefore, wireless caching is an appealing option for supporting cellular VoD streaming while providing a capacity enhancement in both the RAN and the backhaul. More importantly, caching has introduced a new type of cellular resource, namely, the cache memory, and additional degrees of freedom to enhance the system performance. {\revise{Recently, an advanced network architecture for exploiting distributed cache memories at BSs, referred as Fog-RAN, has been proposed for 5G in \cite{Peng16FogRAN,Park16TWC:FogRAN,Tandon16GC,Tandon17online}.}}

Meanwhile, due to the broadcast nature of wireless transmission, VoD streaming data is vulnerable to potential eavesdroppers such as non-paying subscribers and malicious attackers. Thus, secure video streaming schemes providing both video data protection and streaming QoS guarantees are needed in 5G cellular networks. However, secure data delivery in cache-enabled transmission was not considered until recently. %\cite{Clancy15IT:Limits,AwanICC15:D2D,LandICC16:HetNets}. 
The existing works \cite{Clancy15IT:Limits,AwanICC15:D2D,LandICC16:HetNets} were motivated by the coded caching scheme proposed in \cite{Niesen14IT:CodedCaching}. Specifically, each user is equipped with a local cache to pre-store parts of a popular video content. By properly encoding the cached and the delivered content (e.g., via index coding), coded multicast delivery opportunities are enabled for serving various user requests at high delivery rates \cite{Niesen14IT:CodedCaching}. In \cite{Clancy15IT:Limits}, a coded caching scheme was proposed to guarantee secure information delivery when eavesdroppers attempt to intercept the video data over a multicast link. To ensure communication secrecy, the cached and the delivered contents are encoded/encrypted using random secret keys and secure coded multicast delivery is enabled based on Shannon's one-time pad method. Coded caching was extended to device-to-device (D2D) cellular networks in~\cite{AwanICC15:D2D}, where a sophisticated key generation and encryption scheme was proposed. However, the encryption methods in \cite{Clancy15IT:Limits,AwanICC15:D2D} can incur significant signaling overhead for sharing the secret keys; for example, one-time pad based methods typically require the size of the secret keys to be as large as the file size. In \cite{LandICC16:HetNets}, a non-encryption based secure video delivery scheme, which prevents eavesdroppers from obtaining the number of coded packets required for successful video file recovery, was proposed for cache-enabled heterogeneous small cell networks. Nevertheless, an information-theoretic characterization of the secrecy considered in~\cite{LandICC16:HetNets} is missing.

On the other hand, MIMO-based physical layer security (PLS) techniques have been proposed to guarantee information-theoretic secrecy in 5G cellular networks \cite{Khisti10IT:MISOME,Khisti10IT:MIMOME,Liu15TWC:SDOF}. Different from one-time pad or encryption methods, PLS techniques opportunistically exploit the inherent properties of wireless channels to enhance communication secrecy without using secret keys. {\revise{In an %$N_{\mathrm{t}}\times N_{\mathrm{r}}$ 
MIMO wiretap channel with full channel state information (CSI), information-theoretic studies have revealed that the maximal number of secure degrees of freedom (s.d.o.f.)\footnote{Strictly positive s.d.o.f. indicate that the system's secrecy capacity can be scaled up by increasing the transmit power.} enabled by multiple antennas is given by $\min([N_{\mathrm{t}}-N_{\mathrm{e}}]^{+}, N_{\mathrm{r}})$ \cite{Khisti10IT:MISOME,Khisti10IT:MIMOME,Liu15TWC:SDOF,Jafar16IT},}} where $N_{\mathrm{t}}$, $N_{\mathrm{r}}$, and $N_{\mathrm{e}}$ are the number of transmit, receive, and eavesdropping antennas, respectively, and $[x]^{+} \triangleq \max(x,0)$.  Because of their superior performance and low overhead compared to encryption based methods, MIMO-based PLS techniques have been widely advocated for secure transmission in cellular networks. However, to the best of the authors' knowledge, despite the increasing interest in secure cache-enabled communication, the benefits of caching for PLS enhancement have not been investigated in the literature yet.

To fill this void, in this paper, we show that caching is an effective method to enhance the PLS of cellular VoD streaming. Specifically, assume that each BS is equipped with a cache. By caching the same video data at different BSs, several BSs can participate in cooperative joint secure transmission of the video data. Correspondingly, by exploiting the large transmit antenna array formed by the cooperating BSs, the s.d.o.f. can be significantly increased and the PLS can be enhanced \cite{Khisti10IT:MISOME}. Meanwhile, as caching reduces the data sharing overhead typically needed for BS cooperation \cite{Gesbert10JSAC:CoMP}, the s.d.o.f. can be increased even if the cellular network has capacity-constrained backhaul links. 
Interestingly, since data caching takes place \emph{a priori}, e.g., in the early mornings when cellular traffic is low, the overhead of data sharing and channel estimation incurred by the cache-enabled joint transmission is comparable to that of traditional coordinated beamforming \cite[Section IV-A]{Gesbert10JSAC:CoMP} adopted for intercell interference mitigation, particularly when the cache capacity is large.

In %the conference version of this work 
\cite{Xiang16:CoMP}, we have investigated cache-enabled cooperative joint secure transmission assuming %a 
perfect knowledge of the channels between the BSs and the legitimate users as well as the channels between the BSs and the eavesdropper. Yet, in practical systems, the channels to passive eavesdroppers are not perfectly known 
since %the eavesdropper is expected to 
an eavesdropper can remain silent for long periods of time. {\revise{This leads to a reduction of the s.d.o.f. for cooperative transmission \cite{Liu15TWC:SDOF,Jafar16IT} and increases the likelihood of data leakage.}} In this paper, we extend \cite{Xiang16:CoMP} to mitigate the impact of imperfect CSI knowledge. To enhance communication secrecy, artificial noise (AN)-based jamming is applied to effectively interfere the eavesdropper's reception \cite{Negi08TWC:AN,LiTSP11:Opt-Robust,Ng15TWC}. In particular, since the AN is generated randomly and locally at the BSs involving neither the cache nor the backhaul, cooperative jamming by all BSs is considered in this paper for achieving high power efficiency. Moreover, to prevent data leakage under imperfect CSI and as power efficiency is of paramount importance for the design of future communication systems, the joint optimization of cooperative transmission and AN-based jamming is formulated as a robust optimization problem for minimization of the total transmit power while providing QoS and guaranteeing communication secrecy during video delivery. %The objective is to provide as-flat-as-possible secrecy rate for video streaming applications while using a low transmit power. 

{\revise{We note that cache-enabled cooperative transmission for transmit power minimization and delivery time minimization has been  investigated in \cite{Liu13TSP:CoMP,Liu14TSP:CoMP,chen16cooperative,Tao16TWC:Multicast} and \cite{Tandon16GC,Tandon17online}, respectively. However, these works did not consider PLS nor backhaul capacity constraints. Hence, the solutions proposed in~\cite{Liu13TSP:CoMP,Liu14TSP:CoMP,chen16cooperative,Tao16TWC:Multicast,Tandon16GC,Tandon17online} are not applicable to the considered problem, which motivates this work.}} The main contributions of this paper are as follows: 
\begin{itemize}
\item We propose caching as a mechanism to enhance secure cellular video streaming with limited backhaul capacity. Thereby, the cache is instrumental for reducing the backhaul traffic and supporting secure data transmission via BS cooperation.

\item Assuming imperfect CSI knowledge, we formulate a two-stage non-convex robust optimization problem for the minimization of the total BS transmit power while satisfying QoS and secrecy constraints for video delivery. Effective caching and delivery algorithms with polynomial-time computational complexity are developed to solve the problem. In particular, we show that the proposed algorithms are asymptotically optimal when the cache capacity and the number of data sets adopted for cache training are sufficiently large.

\item Simulation results show that the proposed schemes can efficiently utilize the cache capacity to enhance PLS and reduce the total BS transmit power. Moreover, the proposed robust algorithms ensure secure communication even for imperfect CSI knowledge. 
\end{itemize}
The remainder of this paper is organized as follows. In Section~\ref{sec:System-Model}, we present the system model for cache enabled cooperative video delivery. The formulation and solution of the proposed two-stage robust control problem are provided in Sections~\ref{sec:Problem-Formulation}~and~\ref{sec:Problem-Solution}, respectively. In Section~\ref{sec:Simulation-Results}, the performance of the proposed robust algorithms is evaluated by simulation, and finally, Section~\ref{sec:Conclusion} concludes the paper.

\emph{Notation:} Throughout this paper, $\mathbb{R}$ and $\mathbb{C}$ denote the sets of real and complex numbers, respectively; $\mathbb{C}^{L}$ denotes the set of $L \times L$ complex matrices; $\mathbf{I}_{L}$, $\mathbf{1}_{L}$, and $\mathbf{0}_{L}$ are the $L\times L$ identity, all-one, and zero matrices, respectively; $\mathrm{diag}(\mathbf{v})$ is a diagonal matrix with the diagonal elements given by the elements of vector $\mathbf{v}$; $(\cdot)^{T}$ and $(\cdot)^{H}$ are the transpose and complex conjugate transpose operators, respectively; $\mathrm{tr}(\cdot)$, $\mathrm{rank}(\cdot)$, $\det(\cdot)$, $\lambda_{\max}(\cdot)$, and $\left\Vert \cdot \right\Vert_F$ denote the trace, rank, determinant, maximum eigenvalue, and Frobenius norm of a square matrix, respectively; $\mathrm{Pr}(\cdot)$ and $\mathbb{E}(\cdot)$ denote the probability mass operator and the expectation operator, respectively; the circularly symmetric complex Gaussian distribution is denoted by $\mathcal{CN} (\boldsymbol{\mu}, \mathbf{C})$ with mean vector $\boldsymbol{\mu}$ and covariance matrix $\mathbf{C}$; $\sim$ stands for ``distributed as"; $|\mathcal{X}|$ and $\mathrm{conv}(\mathcal{X})$ represent the cardinality and the convex hull of set $\mathcal{X}$, respectively; $\mathcal{X}\times\mathcal{Y}$ denotes the Cartesian product of sets $\mathcal{X}$ and $\mathcal{Y}$; $\mathbf{A}\succeq\mathbf{0}$ ($\mathbf{A}\succ\mathbf{0}$) indicates that matrix $\mathbf{A}$ is positive semidefinite (definite); $\nabla_{\mathbf{X}}f\left(\mathbf{X}\right)$ denotes the complex-valued gradient of $f(\mathbf{X})$ with respect to matrix $\mathbf{X}$; finally, $\left\lfloor \cdot\right\rfloor $ denotes the rounding operator %the largest integer smaller than or equal to ...
and $\left(_k^n\right)$ is the binomial coefficient. %$n$-choose-$k$ combination operator. 

\vspace{-0.2cm}
\section{\label{sec:System-Model}System Model}

\begin{table*}[t]
\centering
{\revise{\protect\protect\caption{\label{tab2}List of Key Notations.}
\vspace{-.3cm}
\small{ 
\begin{tabular}{|c|c|}
\hline 
 {$\mathcal{M}$, $\mathcal{K}$, $\mathcal{M}_{f,l}^{\mathrm{Coop}}$}    &  {Sets of $M$ BSs, $K$ LRs, subset of cooperating BSs for delivery of subfile $(f,l)$} \tabularnewline
\hline  
{$\mathcal{F}$, $\mathcal{L}$} & {Sets of $F$ video files and $L$ subfiles per file}\tabularnewline
\hline 
{$\boldsymbol{\rho}\triangleq(k,f,l)$, $\mathcal{S}$} & {Request of LR $k$ for subfile $(f,l)$ and set of user requests }\tabularnewline
\hline 
{$c_{f,l,m}$, $b_{f,l,m}$, $q_{f,l,m}$} & {Caching, backhaul loading, and cooperative delivery decisions for subfile $(f,l)$ at BS $m$}\tabularnewline
\hline 
{$\mathbf{w}_{m,\boldsymbol{\rho}}$, $\mathbf{w}_{\boldsymbol{\rho}}$} & {Beamforming vectors for BS $m$ and BS set $\mathcal{M}$}\tabularnewline
\hline 
{$\mathbf{v}$, $\mathbf{V}$ } & { AN and its covariance matrix}\tabularnewline
\hline 
{$B_m^{\max}$, $C_m^{\max}$ } & { Backhaul link capacity and cache size at BS $m$ }\tabularnewline
\hline 
{$\Gamma_{\boldsymbol{\rho}}$, $R_{\boldsymbol{\rho}}$, $R_{\boldsymbol{\rho}}^{\mathrm{sec}}$ } & {SINR, achievable rate, and achievable secrecy rate at LR ${\boldsymbol{\rho}}$ }\tabularnewline
\hline 
{$R_{\mathrm{e},\boldsymbol{\rho}}$} &  {Capacity of the ER for eavesdropping LR ${\boldsymbol{\rho}}$}
\tabularnewline
\hline 
\end{tabular} }
}}
\end{table*}

\begin{figure*}[t]
\vspace{-0.2cm}
 \centering 
 \subfloat[] {\includegraphics[height=2.8in]{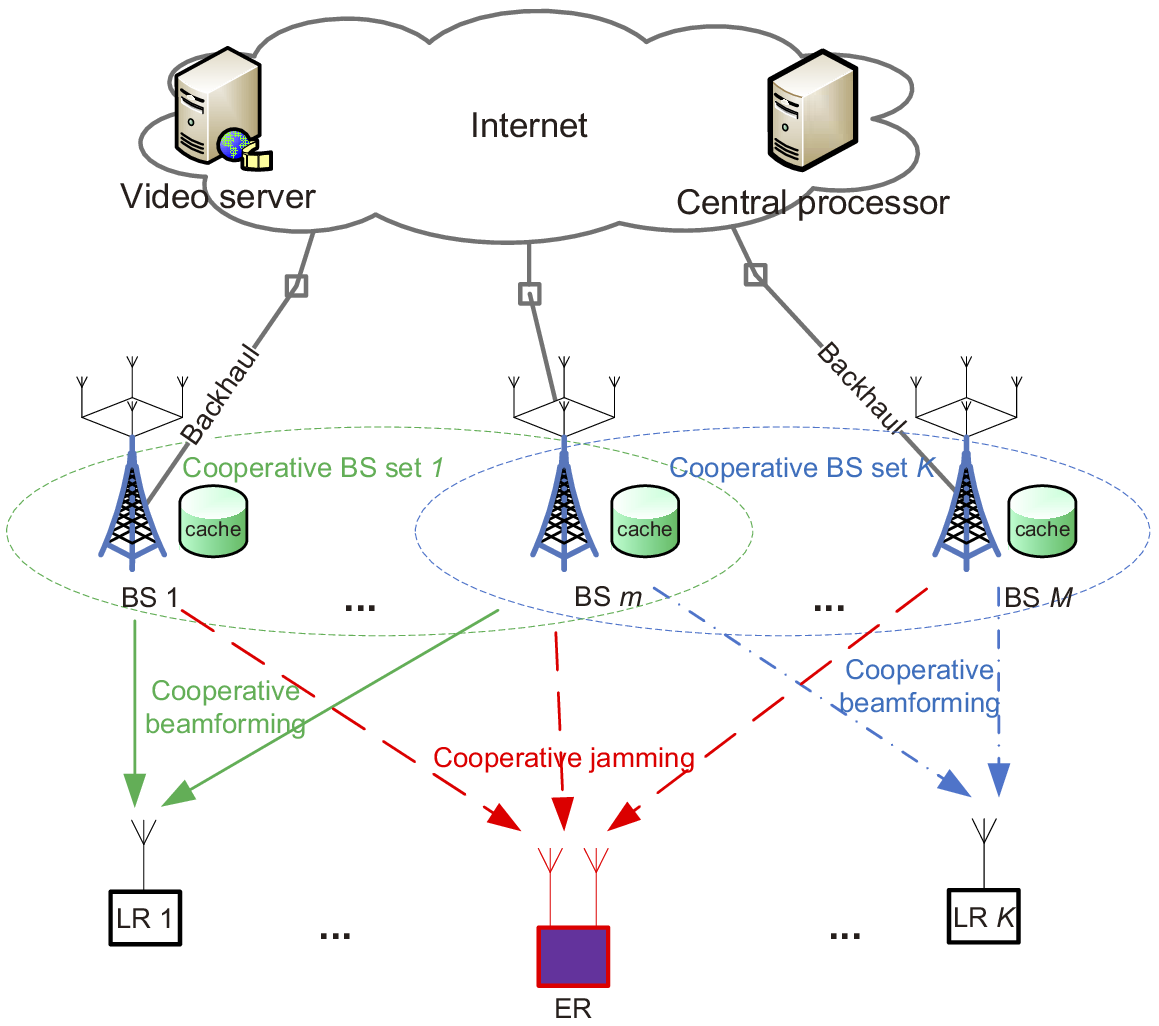} \label{fig:System-Model} } \quad
 \subfloat[] {\includegraphics[height=2.8in]{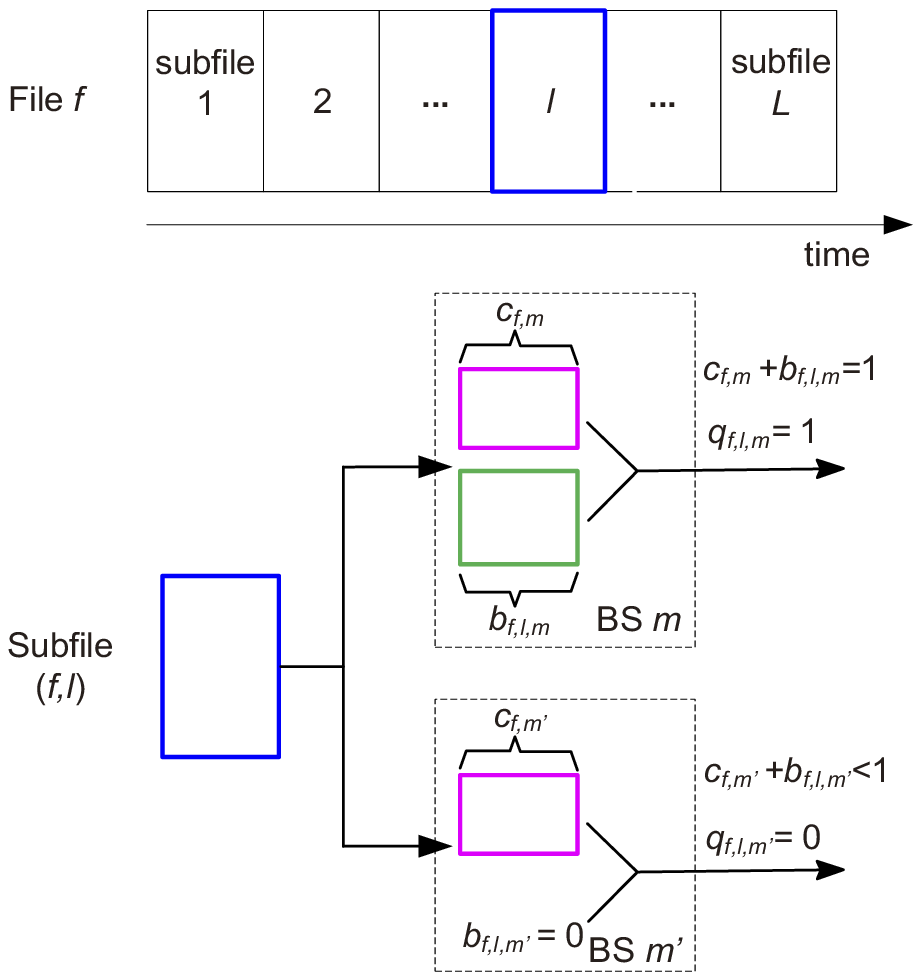}  \label{fig1b} }
\vspace{-0.2cm}
 \caption{{\small (a) System model for cooperative beamforming and jamming for secure video delivery. %E.g., the requested video files of LRs $1$ and $K$ are cached at BS sets $1$ and $K$ for cooperative transmission, respectively; meanwhile, all BSs cooperate for jamming the ER. 
 {\revise{(b) Each file is split/segmented into multiple subfiles/segments. Each subfile can be shared through both caching and backhaul loading to enable cooperative joint transmission.}}
}}
\end{figure*}

We consider video streaming in the downlink of a multi-cell cellular network as shown in Fig.~\ref{fig:System-Model}. A set of BSs, $\mathcal{M}=\{1,\ldots,M\}$, each equipped with $N_{\mathrm{t}}$ antennas, broadcast the video data to a set of single-antenna legitimate receivers (LRs), $\mathcal{K}=\{1,\ldots,K\}$. Since the broadcasted video data may be overheard by a passive eavesdropping receiver (ER), securing the video delivery is desirable to reduce the chance of potential information leakage. We assume that the ER is equipped with $N_{\mathrm{e}}$ antennas\footnote{The ER may represent a set of distributed ERs with a total number of $N_{\mathrm{e}}$ antennas which are connected to perform joint eavesdropping.}.

The video server located at the Internet edge owns a library of video files, $\mathcal{F}=\{1,\ldots,F\}$, which are intended for delivery. The size of video file $f\in \mathcal{F}$ is $V_{f}$ bits. The BSs are connected to the video server via dedicated ``last-mile'' wired backhaul links such as digital subscriber lines. We assume that the backhaul links are secure. However, since each backhaul is shared by different types of traffic (e.g., voice, data, multimedia, control signaling, etc.), the backhaul capacity available for supporting video streaming may be time-varying and limited. To reduce the backhaul traffic, a cache is deployed at each BS for pre-storing the video data.

The cache-enabled system is time-slotted and its operation is divided into two stages. In the first stage, a portion of the video files is cached at the BSs. To reduce the system overhead, the cache is updated when the network utilization is low, %by utilizing vacant network capacity, 
e.g., during  early mornings. %when cellular traffic is low. 
In the second stage, users request video files and in response, a subset of the BSs cooperate  to address the requests. We assume that there is a central processor (CP) also located at the Internet edge, which is capable of performing computationally intensive signal processing tasks. The video caching and delivery control decisions are determined at the CP and conveyed to the BSs via the backhaul links. {\revise{In this paper, we focus on static caching for notational convenience. Nevertheless, the proposed schemes can be extended to spatio-temporally dynamic caching by executing the corresponding algorithms multiple times across space and time.}} {\revise{A list of key notations is shown in Table~\ref{tab2}.}}

\vspace{-0.4cm}
\subsection{Caching and Backhaul Loading in Two Stages}

{\revise{We assume hypertext transfer protocol (HTTP) based video streaming \cite{Ma11:MobVid}. Thereby, file $f$ is split/segmented into $L$ subfiles/segments of equal sizes, i.e., $V_{f}/L$, and each subfile $(f,l) \! \in \! \mathcal{F}\times\mathcal{L}$ is delivered in one time slot\footnote{\revise{Herein, one time slot corresponds to a streaming session in HTTP video streaming. To facilitate a succinct formulation of the transmit power minimization problem for video file delivery, we neglect any inter-session delay caused by packet scheduling and user interruptions.}}, where $L\! \gg \!1$ and $\mathcal{L} \! = \! \{1,\ldots,L\}$, cf. Fig. \ref{fig1b}.}} Let binary variable $q_{f,l,m} \! \in \! \{0,1\}$ indicate the participation of BS $m\in\mathcal{M}$ in the cooperative transmission of subfile $(f,l)$. The set of BSs cooperating for delivering subfile $(f,l)$ is then defined as $\mathcal{M}_{f,l}^{\mathrm{Coop}}\triangleq\left\{ m \! \in \!\mathcal{M}\mid q_{f,l,m}=1\right\} \! \subseteq \! \mathcal{M}$. To facilitate cooperative BS transmission, the video data can be conveyed to the cooperating BSs in two manners: (1) caching the data ahead of time; (2) loading it via the backhaul links instantaneously during delivery. The data caching decisions are determined in the first stage based on the statistics or historical records of the user requests{\revise{\footnote{\revise{Similar to file popularity estimation \cite{Paschos16WC}, the proposed cache training is performed based on profiles of user requests (and CSI) and does not require explicit \emph{a priori} knowledge of the file popularity. Note that the proposed caching scheme is offline. Online cache training based on learning was studied in \cite{ICC14:Learning}.}} }}. The cache status remains unchanged once it is updated according to the caching decisions. Since the information of the user requests, the channel state, and the backhaul capacity are only known online at the time of request, joint optimization of backhaul loading and BS cooperative transmission is deferred to the second stage when this information is (partially) available. We note that two-stage control schemes have also been widely studied in the literature of stochastic control  \cite{Birge2011IntroSP,Shapiro2009LectSP} as they can incorporate online information for improved system performance. For a similar reason, two-stage control is  a popular choice for operating cache-enabled cellular networks \cite{Liu13TSP:CoMP,Liu14TSP:CoMP,Xiang17TVT:CLCaching}.

For the purpose of data sharing, we cache $c_{f,l,m}\in[0,1]$ portion and load via the backhaul $b_{f,l,m}\in[0,1]$ portion of subfile $(f,l)$ at BS $m$, respectively. We assume that the same portion of each subfile of file $f$ is cached at BS $m$, i.e., $c_{f,l,m} = c_{f,m}$, $\forall l \in \mathcal{L}$. Fig.~\ref{fig1b} illustrates the caching and backhaul loading process for cooperative transmission of subfile $(f,l)$.  We can establish the following relation between caching, backhaul loading, and cooperation formation 
\vspace{-.2cm}
\begin{equation}
\begin{array}{c}
b_{f,l,m}=(1-c_{f,m})q_{f,l,m}. 
\end{array}
\label{eq1-coop}
\end{equation}
That is, successful cooperative transmission of subfile $(f,l)$ is possible, i.e., $q_{f,l,m}=1$, only when its data symbols are fully available at BS $m$ such that $b_{f,l,m}+c_{f,m}=1$. Otherwise, \eqref{eq1-coop} enforces $q_{f,l,m}=0$ and $b_{f,l,m}=0$.

The aggregate backhaul load of BS $m\in\mathcal{M}$ is given by
\vspace{-.2cm}
\begin{equation}
B_{m,l} = \sum\nolimits _{f\in\mathcal{F}}b_{f,l,m}Q_{f},
\end{equation}
where the fixed parameter $Q_{f}$ (in bps) represents the data rate required to load subfile $(f,l)$ via the backhaul at BS $m$. We have $Q_{f}=V_{f}/(\tau L)$ or equivalently $\frac{c_{f,m}V_{f}}{L}+b_{f,l,m}Q_{f}\tau=\frac{V_{f}}{L}$, where $\tau$ denotes the duration of a time slot. Note that $B_{m,l} \le \sum\nolimits _{f\in\mathcal{F}}q_{f,l,m}Q_{f}$, in other words, caching reduces the backhaul capacity required for cooperative BS transmission.  

\vspace{-0.4cm}
\subsection{Cooperative Beamforming and Jamming for Secure Video Delivery}
When subfile $(f,l)$ is available at BS subset $\mathcal{M}_{f,l}^{\mathrm{Coop}}$, these BSs employ joint cooperative beamforming and jamming to deliver the subfile. Assume that an LR requests one (sub)file at a time%
\footnote{If an LR requests several files simultaneously, orthogonal frequency multiple access (OFDMA) can be adopted to simultaneously deliver multiple video data streams to a single-antenna LR \cite{Tse2005Fundamentals}. For a given subcarrier assignment, the model considered in this paper still applies at the expense of extra notations. 
} \cite{Liu14TSP:CoMP}. We denote the request of LR $k$ for subfile $(f,l)$ by $\boldsymbol{\rho}\triangleq(k,f,l)$ and the set of user requests by $\mathcal{S}\subseteq\mathcal{K}\times\mathcal{F}\times\mathcal{L}$. Here, $\mathcal{S}$ is known at the beginning of the online delivery stage. With a slight abuse of notation, in the following, the LRs are also indexed by $\boldsymbol{\rho}$ when the requested (sub)files need to be specified. Note that there is a one-to-one correspondence between $\boldsymbol{\rho}$ and $k$.

{\revise{The data symbols of subfile $(f,l)$ for serving request $\boldsymbol{\rho}$ are denoted by $s_{\boldsymbol{\rho}}\in\mathbb{C}$ and are modeled as complex Gaussian random variables\footnote{{\revise{Gaussian data symbols are capacity-achieving for the additive white Gaussian noise channel, and hence, are commonly assumed for analytical formulation of the achievable rate \cite{Tse2005Fundamentals,Cover12IT}. On the other hand, if the modulation and coding schemes (MCS) employed in practical VoD systems are not capacity-achieving, the achievable rate expression can be amended by introducing an equivalent signal-to-noise ratio (SNR), $\frac{SNR}{\Delta}$, where $\Delta\ge1$ defines the SNR gap between the channel capacity and the rate achievable with the adopted MCS \cite{Cioffi91MC}.}}} with $s_{\boldsymbol{\rho}}\sim\mathcal{CN}(0,\,1)$ \cite[Ch. 5]{Tse2005Fundamentals}, \cite[Ch. 9]{Cover12IT}.}} Let $\mathbf{w}_{m,\boldsymbol{\rho}}\in\mathbb{C}^{N_{\mathrm{t}}\times1}$ be the beamforming vector used at BS $m\in\mathcal{M}$ for sending symbol $s_{\boldsymbol{\rho}}$. As the cooperative BS set $\mathcal{M}_{f,l}^{\mathrm{Coop}}$ dynamically changes with the cache status during backhaul loading, cf. \eqref{eq1-coop},  we impose 
\vspace{-.2cm}
\begin{equation}
 (1-q_{f,l,m})\mathbf{w}_{m,\boldsymbol{\rho}}=\mathbf{0}, \;\; \forall m \in \mathcal{M}, \; \forall \boldsymbol{\rho} \in \mathcal{S},
\label{coopbeam} 
\end{equation}    
to adaptively adjust the beamforming vectors, $\mathbf{w}_{m,\boldsymbol{\rho}}$, according to the cooperation decisions, $q_{f,l,m}$. Based on \eqref{coopbeam}, we have $\mathbf{w}_{m,\boldsymbol{\rho}} \!=\! \mathbf{0}, \forall m \notin \mathcal{M}_{f,l}^{\mathrm{Coop}}$. Without loss of generality, let $\mathbf{w}_{\boldsymbol{\rho}} \!\triangleq\! [\mathbf{w}_{1, \boldsymbol{\rho}}^{H}, \ldots, \mathbf{w}_{M,\boldsymbol{\rho}}^{H} ]^{H} \\ \in \mathbb{C}^{MN_{\mathrm{t}}\times1}$ be the joint beamforming vector for serving request $\boldsymbol{\rho}$. Furthermore, complex Gaussian distributed AN, $\mathbf{v}\in\mathbb{C}^{MN_{\mathrm{t}}\times1}$, is sent cooperatively by BS set $\mathcal{M}$ to proactively interfere the reception of the ER. We assume  $\mathbf{v}\sim\mathcal{CN}(\mathbf{0},\,\mathbf{V})$, where $\mathbf{V}$ is the covariance matrix of the artificial noise, i.e., $\mathbf{V}\triangleq\mathbb{E}\left[\mathbf{v}\mathbf{v}^{H}\right]$. The joint transmit signal of BS set $\mathcal{M}$, denoted as $\mathbf{x}\in\mathbb{C}^{MN_{\mathrm{t}}\times1}$, is thus given by 
\vspace{-.2cm}
\begin{equation}
\mathbf{x}=\sum\nolimits _{\boldsymbol{\rho}\in\mathcal{S}}\mathbf{w}_{\boldsymbol{\rho}}{s}_{\boldsymbol{\rho}}+\mathbf{v}.
\label{txsignal}
\end{equation}  
Note that \eqref{coopbeam} and \eqref{txsignal} describe a flexible BS cooperation topology, which allows any cooperative set $\mathcal{M}_{f,l}^{\mathrm{Coop}} \subseteq \mathcal{M}$,  provided that $\mathcal{M}_{f,l}^{\mathrm{Coop}}$ ensures problem feasibility. For example, 
\eqref{coopbeam}  and \eqref{txsignal} correspond to joint transmission with full BS cooperation when $|\mathcal{M}_{f,l}^{\mathrm{Coop}}| = M, \forall f$, and coordinated beamforming when $|\mathcal{M}_{f,l}^{\mathrm{Coop}}| = 1, \forall f$, respectively. 

\vspace{-0.4cm}
\subsection{Channel Model and Channel State Information}  

We assume a frequency flat fading channel model. The received signals at LR $\boldsymbol{\rho}\in\mathcal{S}$ and the ER, denoted by $y_{\boldsymbol{\rho}}\in\mathbb{C}$ and $\mathbf{y}_{\mathrm{e}}\in\mathbb{C}^{N_{\mathrm{e}}\times1}$, respectively, are given by 
\begin{alignat}{1}
y_{\boldsymbol{\rho}} & =\mathbf{h}_{\boldsymbol{\rho}}^{H}\mathbf{x}+z_{\boldsymbol{\rho}} \label{eq:channel}
 \\
& =\underset{\textrm{desired signal}}{\underbrace{\mathbf{h}_{\boldsymbol{\rho}}^{H}\mathbf{w}_{\boldsymbol{\rho}} s_{\boldsymbol{\rho}}} } + \underset{\textrm{multiuser interference}}{ 
   \underbrace{\sum\nolimits_{\boldsymbol{\rho}'\in\mathcal{S},\boldsymbol{\rho}' \neq \boldsymbol{\rho}}\mathbf{h}_{\boldsymbol{\rho}}^{H}\mathbf{w}_{\boldsymbol{\rho}'} s_{\boldsymbol{\rho}'}} } +  \underset{\textrm{injected AN }}{ \underbrace{ \mathbf{h}_{\boldsymbol{\rho}}^{H}\mathbf{v} }} +z_{\boldsymbol{\rho}} , \nonumber %\, \boldsymbol{\rho}\in\mathcal{S},  
\end{alignat}
and $\mathbf{y}_{\mathrm{e}}=\mathbf{G}^{H}\mathbf{x}+\mathbf{z}_{\mathrm{e}}$, where 
\begin{alignat}{1}
\mathbf{h}_{\boldsymbol{\rho}} & =[\mathbf{h}_{1,\boldsymbol{\rho}}^{H},\ldots,\mathbf{h}_{M,\boldsymbol{\rho}}^{H}]^{H}\in\mathbb{C}^{MN_{\mathrm{t}}\times1}  \quad \textrm{and } \\
\mathbf{G} &=[\mathbf{G}_{1}^{H},\ldots,\mathbf{G}_{M}^{H}]^{H}\in\mathbb{C}^{MN_{\mathrm{t}}\times N_{\mathrm{e}}} \nonumber
\end{alignat}
are the channel vectors/matrices from BS set $\mathcal{M}$ to LR $\boldsymbol{\rho}$ and the ER, respectively. $\mathbf{h}_{m,\boldsymbol{\rho}}\in\mathbb{C}^{N_{\mathrm{t}}\times1}$ and $\mathbf{G}_{m}\in\mathbb{C}^{N_{\mathrm{t}}\times N_{\mathrm{e}}}$ model the channels between BS $m\in\mathcal{M}$ and the corresponding LR/ER receivers; and $z\mathbf{_{\boldsymbol{\rho}}}\sim\mathcal{CN}(0,\sigma^{2})$ and $\mathbf{z}_{\mathrm{e}}\sim\mathcal{CN}(\mathbf{0},\sigma_{\mathrm{e}}^{2}\mathbf{I}_{N_{\mathrm{e}}})$ are the zero-mean complex Gaussian noises at the LRs and the ER with variance $\sigma^{2}$ and covariance matrix $\sigma_{\mathrm{e}}^{2}\mathbf{I}_{N_{\mathrm{e}}}$, respectively. Here, the channel vectors/matrices capture the joint effects of multipath fading and path loss; in addition, $z_{\boldsymbol{\rho}}$ and $\mathbf{z}_{\mathrm{e}}$ represent the joint effects of thermal noise and possible out-of-system interference.

For channel estimation, we assume a time division duplex (TDD) system with slowly time-varying channels. At the beginning of each time slot, the LRs send orthogonal pilot sequences simultaneously in the uplink. Due to channel reciprocity, the LRs' downlink channels are estimated by measuring the pilot signals. We assume that the pilot sequences are long enough such that the CSI of the LRs can be reliably estimated at the CP{\revise{\footnote{\revise{In practice, the CSI of the LRs can be imperfect if channel estimation errors occur, which degrades communication secrecy. The proposed robust optimization framework can be easily extended to tackle CSI imperfection at the LRs \cite{LiTSP11:Opt-Robust,Ng15TWC}. However, as the imperfect CSI for the ER usually dominates other source of imperfection, we neglect the effect of imperfect CSI at the LRs to keep the paper readable. The results of this paper provide a performance upper bound for the case of imperfect CSI at LRs.}}}}. In contrast, we consider imperfect CSI for the ER\footnote{Even though the ER is passive and silent, it may still be possible to estimate the CSI of the ER based on the local oscillator power that is unintentionally leaked from its radio frequency (RF) front end during eavesdropping  \cite{mukherjee12detecting}.} since the ER does not directly interact with the BSs during channel estimation, which prevents a timely and accurate estimation of its CSI. 
Let $\widehat{\mathbf{G}}\in\mathbb{C}^{MN_{\mathrm{t}}\times N_{\mathrm{e}}}$ be the estimate of the ER channel matrix $\mathbf{G}$. We write $\mathbf{G} = \widehat{\mathbf{G}} + \Delta\mathbf{G}$, where $\Delta\mathbf{G}$ is the estimation error matrix and its value is unknown at the CP. To capture the effect of imperfect CSI, we assume that $\Delta\mathbf{G}$ lies in a continuous set of possible values given by
\vspace{-0.1cm}
\begin{alignat}{1}
\mathcal{U}_{\textrm{e}} \triangleq\left\{ \Delta\mathbf{G}\in\mathbb{C}^{MN_{\mathrm{t}}\times N_{\mathrm{e}}}\mid \left\Vert \Delta\mathbf{G}\right\Vert _{F}^2 \le \varepsilon_{\textrm{e}}^2 \right\}. \label{eq:CSI-ER} 
\end{alignat}
Here, $\mathcal{U}_{\textrm{e}}$ is referred to as the uncertainty set associated with channel estimate $\widehat{\mathbf{G}}$ and $\varepsilon_{\textrm{e}} \ge 0$ is a measure for the accuracy of the channel estimate of the ER. In practice, the value of $\varepsilon_{\textrm{e}}$ depends on the channel coherence time and the adopted channel estimation method. {\revise{Note that the above imperfect CSI model is commonly adopted in the literature \cite{LiTSP11:Opt-Robust,Ng15TWC}, and the special case $\varepsilon_{\textrm{e}} = 0$ corresponds to the optimistic scenario when $\mathbf{G}$ is perfectly known at the CP.}} 

\subsection{Achievable Secrecy Rate}

The achievable rate at LR $\boldsymbol{\rho} \in \mathcal{S}$, denoted by $R_{\boldsymbol{\rho}}$, is given by 
\begin{alignat}{1}
R_{\boldsymbol{\rho}}  &= \log\left(1+\Gamma_{\boldsymbol{\rho}}\right), \label{eq:LR-rate} \\
\qquad \Gamma_{\boldsymbol{\rho}}  &=\frac{\frac{1}{\sigma^{2}}\left|\mathbf{h}_{\boldsymbol{\rho}}^{H}\mathbf{w}_{\boldsymbol{\rho}}\right|^{2}}{1+\frac{1}{\sigma^{2}}\sum_{\boldsymbol{\rho}'\in\mathcal{S},\boldsymbol{\rho}'\neq\boldsymbol{\rho}}\left|\mathbf{h}_{\boldsymbol{\rho}}^{H}\mathbf{w}_{\boldsymbol{\rho}'}\right|^{2}+\frac{1}{\sigma^{2}}\mathbf{h}_{\boldsymbol{\rho}}^{H}\mathbf{V}\mathbf{h}_{\boldsymbol{\rho}}}, \nonumber %\quad\boldsymbol{\rho}\in\mathcal{S},
\end{alignat}
where $\Gamma_{\boldsymbol{\rho}}$ is the received signal-to-interference-plus-noise ratio (SINR) at LR $\boldsymbol{\rho}$. In \eqref{eq:LR-rate}, the term $\sum_{\boldsymbol{\rho}'\in\mathcal{S},\boldsymbol{\rho}'\neq\boldsymbol{\rho}}\left|\mathbf{h}_{\boldsymbol{\rho}}^{H}\mathbf{w}_{\boldsymbol{\rho}'}\right|^{2}$ accounts for the interference power caused by multiuser transmission, cf.~\eqref{eq:channel}.

To guarantee secure VoD streaming, the proposed secure delivery scheme is designed to avoid VoD data leakage even under worst-case conditions. Specifically, we assume that the ER can eavesdrop the information intended for LR $\boldsymbol{\rho}$ after canceling the interference caused by all other LRs. This is possible if the ER adopts advanced receiver structures such as successive interference cancellation decoders \cite[Ch. 8.3]{Tse2005Fundamentals}. Thus, the achievable secrecy rate for LR $\boldsymbol{\rho}$ is modeled as \cite{Khisti10IT:MISOME,LiTSP11:Opt-Robust},
\vspace{-.2cm}
\begin{equation}
\begin{aligned}R_{\boldsymbol{\rho}}^{\mathrm{sec}} & =\left[R_{\boldsymbol{\rho}}-R_{\mathrm{e},\boldsymbol{\rho}}\right]^{+},\;\boldsymbol{\rho}\in\mathcal{S},\end{aligned}
\label{eq:sec-rate}
\end{equation}
where $R_{\mathrm{e},\boldsymbol{\rho}}$ denotes the capacity of the ER for decoding subfile $(f,l)$ of LR $\boldsymbol{\rho}$ and is given by \cite[Chapter 8]{Tse2005Fundamentals} 
\vspace{-0.2cm}
\begin{equation}
R_{\mathrm{e},\boldsymbol{\rho}}  =\log\det\Big(\mathbf{I}_{N_{\textrm{e}}}+\frac{1}{\sigma_{\textrm{e}}^{2}}\mathbf{G}\mathbf{Z}_{\mathrm{e},\boldsymbol{\rho}}^{-1}\mathbf{G}^{H}\mathbf{w}_{\boldsymbol{\rho}}\mathbf{w}_{\boldsymbol{\rho}}^{H}\Big),\,\boldsymbol{\rho}\in\mathcal{S},
\label{eq:ER-rate}
\end{equation}
with $\mathbf{Z}_{\mathrm{e},\boldsymbol{\rho}}=\mathbf{I}_{N_{\textrm{e}}}+\frac{1}{\sigma_{\textrm{e}}^{2}}\mathbf{G}^{H}\mathbf{V}\mathbf{G}$.

%According to \eqref{eq:sec-rate}, for improving the secrecy rate, the cooperative beamforming and jamming should enlarge the rate difference between $R_{\boldsymbol{\rho}}$ and $R_{\mathrm{e},\boldsymbol{\rho}}$. This means that multiuser interference should be reduced through cooperative beamforming utilizing a virtual transmit antenna array whose size, according to  \eqref{eq1-coop} and \eqref{coopbeam}, depends on the caching decisions. At the same time, the covariance matrix $\mathbf{V}$ for cooperative jamming has to be also properly designed to  degrade the capacity of the ER channels, cf. \eqref{eq:ER-rate}, while avoiding possible rate loss for the LRs, cf. \eqref{eq:LR-rate}. Therefore, caching, cooperative beamforming, and AN-based jamming have to be judiciously and jointly optimized to reap the cache-enabled secrecy benefits. 

According to \eqref{eq:sec-rate}, for power-efficient secure video delivery, the cooperative beamforming and jamming has to ensure a certain rate difference between $R_{\boldsymbol{\rho}}$ and $R_{\mathrm{e},\boldsymbol{\rho}}$ while consuming as little transmit power as possible. This means that multiuser interference should be reduced through cooperative beamforming across a virtual transmit antenna array whose size, according to  \eqref{eq1-coop} and \eqref{coopbeam}, depends on the caching decisions. At the same time, the covariance matrix $\mathbf{V}$ for cooperative jamming has to be properly designed to  degrade the capacity of the ER channel, cf. \eqref{eq:ER-rate}, without causing interference to the LRs, cf. \eqref{eq:LR-rate}. Therefore, caching, cooperative beamforming, and AN-based jamming have to be judiciously and jointly optimized to reap the benefits of cache-enabled secrecy. 

\vspace{-0.4cm}
\section{\label{sec:Problem-Formulation}Robust Two-Stage Problem Formulation}

In this section, we formulate a two-stage robust optimization problem for minimization of the total BS transmit power while taking into account the secrecy/QoS constraints of the LRs and the imperfect CSI knowledge regarding the ER. In the first stage, the cached video data is optimized \emph{offline}. In the second stage, the cooperative transmission strategy is optimized \emph{online} for a given cache and backhaul status. For ease of discussion, we first present the formulation of the second-stage problem. 

\vspace{-0.4cm}
\subsection{Second-Stage Online Delivery Control }

The BS cooperation formation policy $\mathbf{D}_{\text{\mbox{II}},1}\triangleq[q_{f,l,m},\, b_{f,l,m}]$ and the cooperative transmission policy $\mathbf{D}_{\text{\mbox{II}},2}\triangleq[\mathbf{w}_{\boldsymbol{\rho}},\mathbf{V}]$ (including beamforming and jamming) are optimized in the second stage. For this purpose, we assume that the set of user requests $\mathcal{S}$ is given and the cache status $\left\{ c_{f,m}\right\} $ has already been determined in the first stage. Let $\mathbf{D}_{\text{\mbox{II}}}\triangleq[\mathbf{D}_{\text{\mbox{II}},1},\mathbf{D}_{\text{\mbox{II}},2}]$ be the optimization space of the second (delivery) stage. The considered optimization problem is formulated in \eqref{eq:R1} at the top of the next page, %as follows, %\footnote{Both the first- and the second-stage problems (i.e., R0 and Q0) are defined per time slot, where the time index is dropped for simplicity of notation.}, 
%\vspace{-.2cm}
\begin{figure*}
\begin{align}
\textrm{R0:}\;\minimize%_{\mathbf{D}_{\text{\mbox{II}}}}
\quad & f_{\mathrm{\text{\mbox{II}}}}\triangleq\sum\nolimits _{\boldsymbol{\rho}\in\mathcal{S}}\mathrm{tr}(\mathbf{w}_{\boldsymbol{\rho}}\mathbf{w}_{\boldsymbol{\rho}}^{H}+\mathbf{V})\label{eq:R1}\\
{{\st \quad}} & \textrm{C1: } b_{f,l,m}=(1-c_{f,m})q_{f,l,m}, \; (f,l)\in\mathcal{F}\times\mathcal{L}, \, m\in \mathcal{M},\nonumber \\
&\textrm{C2: } b_{f,l,m}\in[0,1],\; q_{f,l,m}\in\left\{ 0,1\right\},  \; (f,l)\in\mathcal{F}\times\mathcal{L}, \, m\in \mathcal{M}, \nonumber \\
& \textrm{C3: }  \sum\nolimits _{f\in\mathcal{F}}b_{f,l,m}Q_{f}\le B_{m}^{\max},\; \forall l , \forall m, 
\;\;\;\; \textrm{C4:}\; \left( 1- q_{f,l,m} \right) \mathbf{w}_{\boldsymbol{\rho}} = \mathbf{0},\; \forall m, \forall \boldsymbol{\rho}, \nonumber \\
& \textrm{C5: }  \mathrm{tr}\left(\boldsymbol{\Lambda}_{m}\left(\sum\nolimits _{\boldsymbol{\rho}\in\mathcal{S}}\mathbf{w}_{\boldsymbol{\rho}}\mathbf{w}_{\boldsymbol{\rho}}^{H}+\mathbf{V}\right)\right)\le P_{m}^{\max},\quad m\in\mathcal{M}, \nonumber \\ 
& \textrm{C6: } R_{\boldsymbol{\rho}}\ge R_{\boldsymbol{\rho}}^{\textrm{req}},\;\boldsymbol{\rho}\in\mathcal{S}, 
\qquad\qquad  \textrm{C7: }\max_{\Delta\mathbf{G}\in\mathcal{U}_{\textrm{e}}}R_{\mathrm{e},\boldsymbol{\rho}}\le R_{\mathrm{e},\boldsymbol{\rho}}^{\mathrm{tol}},\;\boldsymbol{\rho}\in\mathcal{S},\nonumber \\
{\revise{\var}} \quad & {\revise{ {\mathbf{D}_{\text{\mbox{II}}}} = [q_{f,l,m},\, b_{f,l,m}, \mathbf{w}_{\boldsymbol{\rho}},\mathbf{V}],}} \nonumber
\end{align}
\hrulefill
\end{figure*}
where $B_{m}^{\max}$ is the backhaul capacity available for video sharing at BS $m$, $P_{m}^{\max}$ is the maximum transmit power at BS $m$, and $\boldsymbol{\Lambda}_{m}$ is an $MN_{\mathrm{t}}\times MN_{\mathrm{t}}$ diagonal matrix defined as %
$\boldsymbol{\Lambda}_{m}=\mathrm{diag}(\mathbf{0}_{(m-1)N_{\textrm{t}}\times1}^{T},\mathbf{1}_{N_{\textrm{t}}\times1}^{T},\mathbf{0}_{(M-m)N_{\textrm{t}}\times1}^{T})$, such that $\mathrm{tr}\left(\mathbf{w}_{m,\boldsymbol{\rho}} \mathbf{w}_{m,\boldsymbol{\rho}}^{H}\right)=\mathrm{tr}\left(\boldsymbol{\Lambda}_{m} \mathbf{w}_{\boldsymbol{\rho}} \mathbf{w}_{\boldsymbol{\rho}}^{H}\right)$ holds. By constraints C1, C2, and C4, BS $m$ is only allowed to cooperate in transmitting subfile $(f,l)$ if $q_{f,l,m}=1$ or $b_{f,l,m} + c_{f,l,m} =1$; otherwise, if $q_{f,l,m}=0$ and $b_{f,l,m} + c_{f,l,m} <1$, BS $m$ cannot cooperate as $\mathbf{w}_{\boldsymbol{\rho}} = \mathbf{0}$. Constraints C3 and C5 limit the maximum backhaul rate and the maximum transmit power of each BS, respectively. C6 guarantees a minimum required video delivery rate, $R_{\boldsymbol{\rho}}^{\textrm{req}}$, to provide video streaming QoS for LR $\boldsymbol{\rho}$. C7 ensures that the capacity of the ER is kept below a maximum tolerable secrecy threshold $R_{\mathrm{e},\boldsymbol{\rho}}^{\mathrm{tol}}$ for video data protection. Due to the imperfect CSI of the ER, the capacity of the ER is constrained for all possible estimation error matrices within the uncertainty set, i.e., for any $\Delta\mathbf{G}\in\mathcal{U}_{\textrm{e}}$, in C7, to provide robustness in secure communication. C6 and C7 together guarantee a minimum achievable secrecy rate of $R_{\boldsymbol{\rho}}^{\mathrm{sec}}= [ R_{\boldsymbol{\rho}}^{\textrm{req}} - R_{\mathrm{e},\boldsymbol{\rho}}^{\mathrm{tol}}]^{+}$ for LR $\boldsymbol{\rho}$, provided that problem R0 is feasible.

\begin{rem}
If, instead of \eqref{eq:CSI-ER}, the probability distribution of the channel estimation error matrix is known, e.g., if $\Delta\mathbf{G}$ is complex-valued Gaussian distributed, the robust design in R0 is still applicable and can ensure a probabilistic secrecy outage constraint, e.g., $\Pr(R_{\boldsymbol{\rho}}^{\mathrm{sec}}<[R_{\boldsymbol{\rho}}^{\textrm{req}}-R_{\mathrm{e},\boldsymbol{\rho}}^{\mathrm{tol}}]^{+}) \le \alpha$, by following a similar approach as in \cite{LiTSP11:Opt-Robust}. Meanwhile, if perfect CSI knowledge of the ER is available, i.e., $\varepsilon_{\textrm{e}}=0$ in C7, the optimal solution of R0 provides a performance upper bound for the case of imperfect CSI. However, the methods for solving R0 for $\varepsilon_{\textrm{e}}=0$ and $\varepsilon_{\textrm{e}}>0$ differ slightly in how C7 is dealt with, cf. Section \ref{sec:Problem-Solution}.
\end{rem}

\vspace{-0.4cm}
\subsection{First-Stage Offline Cache Training}

A historical data driven approach~\cite{Xiang17TVT:CLCaching,Birge2011IntroSP} is adopted for optimizing the offline caching in the first stage. Assume that $\Omega$ sets of typical scenario data are available for training the cache, e.g., obtained from past system records. Each set of scenario data consists of the user requests, CSI, and the available backhaul capacities at a particular time instant. The scenario data is indexed by $\omega\in\left\{ 1,\ldots,\Omega\right\} $. Let $\mathbf{C}_{\text{\mbox{I}}}\triangleq[c_{f,m},\,\mathbf{D}_{\text{\mbox{I}},\omega}]$ be the first-stage (caching) optimization space, where $\mathbf{D}_{\text{\mbox{I}},\omega}\triangleq[q_{f,l,m,\omega},b_{f,l,m,\omega},\mathbf{w}_{\boldsymbol{\rho},\omega},\mathbf{V}_{\omega}]$ denotes the auxiliary delivery decisions for scenario $\omega$ during training. We define the feasible delivery set for scenario $\omega$ by $\mathbf{\mathcal{D}}_{\text{\mbox{I}},\omega}\triangleq\left\{ \mathbf{D}_{\text{\mbox{I}},\omega}\mid\textrm{C1, C2, C4--C7}\right\} $, where C1, C2, and C4--C7 need to be reformulated with an augmented system state space. For example, C1 and C2 are rewritten as 
\vspace{-.2cm}
\begin{alignat}{1}
&\textrm{C1: }  b_{f,l,m,\omega}=(1-c_{f,m})  q_{f,l,m,\omega}, \label{C1-2} \\
&\textrm{C2: }  c_{f,m}, b_{f,l,m,\omega} \in [0,1], \; 
q_{f,l,m,\omega} \in \left\{ 0,1\right\}, \nonumber
\end{alignat}
and C4--C7 are similarly formulated.

The objective of the first-stage problem is the minimization of the average transmit power for the considered scenarios, i.e., 
\vspace{-0.2cm}
\begin{align}
\textrm{Q0:}\; \minimize%_{\mathbf{C}_{\text{\mbox{I}}}}
\; & \frac{1}{\Omega}\sum_{\omega=1}^{\Omega}\; f_{\mathrm{\text{\mbox{I}}},\omega}\\
{\st}\; & \ensuremath{\overline{\textrm{C3}}}\textrm{: } \frac{1}{\Omega} \!\! \sum_{\omega=1}^{\Omega} \!\! \sum_{f\in\mathcal{F}}b_{f,l,m,\omega}Q_{f}  \! \le \! \frac{1}{\Omega} \!\! \sum_{\omega=1}^{\Omega}B_{m,\omega}^{\max}, \forall m, \nonumber \\ %\, m\in\mathcal{M},
& \textrm{C8: }  \sum_{f\in\mathcal{F}}c_{f,m}V_{f}\le C_{m}^{\max},\; m\in\mathcal{M}, \nonumber \\
& \textrm{C9: } \mathbf{D}_{\text{\mbox{I}},\omega}\in\mathbf{\mathcal{D}}_{\text{\mbox{I}},\omega},\;\omega\in\left\{ 1,\ldots,\Omega\right\}, \nonumber \\
{\revise{\var}} \; & {\revise{ \mathbf{C}_{\text{\mbox{I}}} = [c_{f,m},\,q_{f,l,m,\omega},b_{f,l,m,\omega},\mathbf{w}_{\boldsymbol{\rho},\omega},\mathbf{V}_{\omega}], }} \nonumber
\end{align}
where $f_{\text{\mbox{I}},\omega}\triangleq\sum_{\boldsymbol{\rho}\in\mathcal{S}}\mathrm{tr}(\mathbf{w}_{\boldsymbol{\rho},\omega}\mathbf{w}_{\boldsymbol{\rho},\omega}^{H}+\mathbf{V}_{\omega})$ is the instantaneous transmit power for scenario $\omega$ and $C_{m}^{\max}$ is the cache capacity at BS $m$. $\ensuremath{\overline{\textrm{C3}}}$ is an average backhaul capacity constraint and C8 is the cache capacity constraint. Considering $\ensuremath{\overline{\textrm{C3}}}$ and C8, the number of cooperating BSs during online VoD streaming depends largely on the values of $C_{m}^{\max}$ and $B_{m,\omega}^{\max}$. Note that $\ensuremath{\overline{\textrm{C3}}}$ is a relaxation of the per-scenario backhaul capacity constraint $\sum_{f\in\mathcal{F}}b_{f,l,m,\omega}Q_{f}\le B_{m,\omega}^{\max}$, $\omega\in\left\{ 1,\ldots,\Omega\right\} $. In the caching phase, when the actual backhaul capacity at the time of delivery is uncertain, $\ensuremath{\overline{\textrm{C3}}}$ avoids the conservative use of the backhaul links, considering that the actual cooperative transmission decisions are deferred to the second stage when the available backhaul capacity is known. Furthermore, $\ensuremath{\overline{\textrm{C3}}}$ facilitates the design of low-complexity asymptotically optimal cache training as will be revealed in Section \ref{sec4-3}.

Problems R0 and Q0 are non-convex mixed-integer nonlinear programs (MINLPs)\footnote{For a non-convex MINLP, even if the integer constraints are relaxed to convex constraints, the problem remains non-convex~\cite{Floudas1995MINLP}.} due to non-convex constraints C6, C7, and the binary optimization variables $q_{f,l,m}\in\left\{ 0,1\right\} $ and $q_{f,l,m,\omega}\in\left\{ 0,1\right\} $. {\revise{Moreover, problems R0 and Q0 involve bilinear constraints C1 and C4.}} Even setting aside secrecy, the formulated two-stage control problem is more practical compared to the problems considered in \cite{Liu14TSP:CoMP,chen16cooperative,Tao16TWC:Multicast}, as it accounts for binary cooperation formation decisions, capacity-constrained backhaul links, and imperfect CSI. However, MINLPs are generally NP-hard and there are no known polynomial-time algorithms to solve them optimally \cite{Floudas1995MINLP}. In Section \ref{sec:Problem-Solution}, to strike a balance between computational complexity and optimality, we present two effective polynomial-time suboptimal algorithms for solving problems R0 and Q0, respectively. We further show that the proposed algorithms are asymptotically optimal when the cache capacity and the number of scenarios considered are sufficiently large, respectively.

\section{\label{sec:Problem-Solution}Solution of Two-Stage Problem}

In this section, the solutions of problems R0 and Q0 are presented. We first tackle problem R0. The approach employed for solving problem R0 is then extended to solve problem Q0. 

\vspace{-0.4cm}
\subsection{\label{sub4-1}Optimal Solution of Problem R0 in Large Cache Capacity Regime}

We first discuss special conditions under which problem R0 is solvable in polynomial time. The corresponding results provide the basis for solving problem R0 in the general case. 

Let $\mathcal{F}(\mathcal{S})\triangleq\left\{ f\mid(k,f,l)\in\mathcal{S}\right\}$ be the set of files requested by the LRs in $\mathcal{S}$, where $\mathcal{F}(\mathcal{S})\subseteq\mathcal{F}$. We define $F(\mathcal{S})\triangleq\left|\mathcal{F}(\mathcal{S})\right|$, which satisfies $F(\mathcal{S})\le\min\left\{ \left|\mathcal{S}\right|,F\right\} $. For a given cache status $\left\{ c_{f,m}\right\} $, C1 is an affine equality constraint which enables the elimination of the backhaul loading decision variables $b_{f,l,m}$. As a result, problem R0 is reformulated as 
\vspace{-.2cm}
\begin{align}
\textrm{R0:}\;\minimize%_{\mathbf{D}_{\text{\mbox{II}}}}
\; & f_{\mathrm{\text{\mbox{II}}}}\label{eq:R1-1}\\
{{\st \;}} & \textrm{C4,}\textrm{ C5, C6, C7, }\mathbf{V}\succeq\mathbf{0}, \nonumber \\
& \overline{\textrm{C2}}\textrm{: } q_{f,l,m}\in\left\{ 0,1\right\}, (f,l)\in \mathcal{F}\times\mathcal{L}, m \in\mathcal{M} ,\nonumber \\
 & \qquad  \widetilde{\textrm{C3}}:\sum_{f\in\mathcal{F}(\mathcal{S})}q_{f,l,m}Q_{f,m}\le B_{m}^{\max},\; \forall l, \forall m, \nonumber \\ %l\in\mathcal{L}, m\in\mathcal{M},
{\revise{\var}} \; & {\revise{ {\mathbf{D}_{\text{\mbox{II}}}} = [q_{f,l,m},\, b_{f,l,m}, \mathbf{w}_{\boldsymbol{\rho}},\mathbf{V}],}} \nonumber
\end{align}
where $Q_{f,m}\triangleq (1-c_{f,m})Q_{f}$ is the ``effective'' data rate required  for loading subfile $(f,l)$ into BS $m$ via the backhaul link and hence, constraints $\widetilde{\textrm{C3}}$ and $\textrm{C3}$ are equivalent. We have the following lemma for the BS cooperation formation in \eqref{eq:R1-1}. 
\vspace{-0.2cm} 
\begin{lem}[Monotonicity of R0]
\label{lem:(Monotonicity-of-R0)} \emph{Let $f_{\mathrm{\text{\mbox{II}}}}^{1}$ and $f_{\mathrm{\text{\mbox{II}}}}^{2}$ denote the optimal objective values for given cooperation sets $\mathcal{M}_{f,l}^{\mathrm{Coop,1}}$ and $\mathcal{M}_{f,l}^{\mathrm{Coop,2}}$, respectively. If $\mathcal{M}_{f,l}^{\mathrm{Coop,1}}\subseteq\mathcal{M}_{f,l}^{\mathrm{Coop,2}},\,\forall(f,l)\in\mathcal{F}(\mathcal{S})\times\mathcal{L}$, then $f_{\mathrm{\text{\mbox{II}}}}^{1}\ge f_{\mathrm{\text{\mbox{II}}}}^{2}$ holds.} 
\end{lem}
\vspace{-0.2cm}
\begin{IEEEproof}
Assume that $\mathbf{V}$ is given. If $\mathcal{M}_{f,l}^{\mathrm{Coop},i}$ is adopted for solving problem R0, let $\mathcal{D}_{\boldsymbol{\rho}}^{{i}}$ and $\mathcal{D}_{m,\boldsymbol{\rho}}^{{i}}$
be the resulting feasible sets of $\mathbf{w}_{\boldsymbol{\rho}}$ and $\mathbf{w}_{m,\boldsymbol{\rho}}$, respectively, where $\mathcal{D}_{\boldsymbol{\rho}}^{{i}}=\prod_{m=1}^{M}\mathcal{D}_{m,\boldsymbol{\rho}}^{{i}}$, ${i}=1,2$. Considering constraint C4, we have $\mathbf{0}\in\mathcal{D}_{m,\boldsymbol{\rho}}^{{i}}$
if $m\in\mathcal{M}$, and $\mathcal{D}_{m,\boldsymbol{\rho}}^{{i}}=\{\mathbf{0}\}$ if $m\notin\mathcal{M}_{f,l}^{\mathrm{Coop},i}$. Besides, $\mathcal{D}_{m,\boldsymbol{\rho}}^{\mathrm{1}}=\mathcal{D}_{m,\boldsymbol{\rho}}^{\mathrm{2}}$ if $m\in\mathcal{M}_{f,l}^{\mathrm{Coop,1}}$ and $m\in\mathcal{M}_{f,l}^{\mathrm{Coop,2}}$. Thus, $\mathcal{D}_{\boldsymbol{\rho}}^{\mathrm{1}}\subseteq\mathcal{D}_{\boldsymbol{\rho}}^{\mathrm{2}}$ holds if $\mathcal{M}_{f,l}^{\mathrm{Coop,1}}\subseteq\mathcal{M}_{f,l}^{\mathrm{Coop,2}}$. 
We have $f_{\mathrm{\text{\mbox{II}}}}^{1}\ge f_{\mathrm{\text{\mbox{II}}}}^{2}$ for any given $\mathbf{V}$, since the objective function of R0 for given $\mathbf{V}$ is only a function of $\mathbf{w}_{\boldsymbol{\rho}}$. This completes the proof. 
\end{IEEEproof}
In systems with large cache capacity, the backhaul capacity constraints C3/$\widetilde{\textrm{C3}}$ can be removed and there is no loss of optimality as the cache can effectively offload the backhaul traffic. Besides, fully cooperative transmission with cooperation set $\mathcal{M}_{f,l}^{\mathrm{F-Coop}}=\mathcal{M}$ is feasible and, based on Lemma~\ref{lem:(Monotonicity-of-R0)}, optimal considering $\mathcal{M}_{f,l}^{\mathrm{Coop}}\subseteq\mathcal{M}_{f,l}^{\mathrm{F-Coop}},\forall\mathcal{M}_{f,l}^{\mathrm{Coop}}$. 

Furthermore, the non-convex problem R0 becomes polynomial time solvable. This is in fact due to the more general result that, whenever the cooperation sets are fixed (i.e., the cooperation formation decision $\mathbf{D}_{\text{\mbox{II}},1} = [q_{f,l,m},\, b_{f,l,m}]$ is known \emph{a priori}), the resulting problem, denoted by R0($\mathbf{D}_{\text{\mbox{II}},2}$), 
\vspace{-.2cm}
\begin{align}
\textrm{R0(\ensuremath{\mathbf{D}_{\text{\mbox{II}},2}}): }\;\minimize%_{\ensuremath{\mathbf{D}_{\text{\mbox{II}},2}} }
\quad & f_{\mathrm{\text{\mbox{II}}}}\\
{{\st \quad}} & \textrm{C4, C5, C6, C7, } \mathbf{V}\succeq\mathbf{0},\nonumber \\
{\revise{\var}} \quad & {\revise{ {\mathbf{D}_{\text{\mbox{II}},2}} = [\mathbf{w}_{\boldsymbol{\rho}},\mathbf{V}],}} \nonumber
\end{align}
is polynomial time solvable. %, where $\mathbf{D}_{\text{\mbox{II}},2}= [ \mathbf{w}_{\boldsymbol{\rho}},\mathbf{V}]$ is the set of optimization variables. 
In particular, despite the seemingly non-convex constraints C4, C6, and C7 included in R0($\mathbf{D}_{\text{\mbox{II}},2}$), we will reveal the hidden convexity of R0($\mathbf{D}_{\text{\mbox{II}},2}$) and show that R0($\mathbf{D}_{\text{\mbox{II}},2}$) can be solved optimally and efficiently via an equivalent convex problem. For this purpose, C4, C6, and C7 are first transformed into convex forms below. 

We first apply the big-M reformulation \cite{Floudas1995MINLP} of bilinear constraint C4, which gives
\begin{equation}
\ensuremath{\overline{\textrm{C4}}}\textrm{: }  \mathrm{tr}\left(\boldsymbol{\Lambda}_{m}\mathbf{w}_{\boldsymbol{\rho}}\mathbf{w}_{\boldsymbol{\rho}}^{H}\right)\le q_{f,l,m}P_{m}^{\max},\; m\in\mathcal{M}, \boldsymbol{\rho}\in\mathcal{S}.  \label{eq:coop3}
\end{equation}
We note that $\ensuremath{\overline{\textrm{C4}}}$ is equivalent to C4, as both constraints enforce $\mathbf{w}_{m,\boldsymbol{\rho}}=\mathbf{0}$ whenever BS $m\notin\mathcal{M}_{f,l}^{\mathrm{Coop}}$ cannot participate in the cooperative transmission of subfile $(f,l)$. For example, if $q_{f,l,m}=0$, %or $b_{f,l,m}+c_{f,m}<1$ (cf. C1), 
we have $\mathrm{tr}\left(\boldsymbol{\Lambda}_{m}\mathbf{w}_{\boldsymbol{\rho}}\mathbf{w}_{\boldsymbol{\rho}}^{H}\right)=\left\Vert \mathbf{w}_{m,\boldsymbol{\rho}}\right\Vert _{2}^{2}=0$, which results in $\mathbf{w}_{m,\boldsymbol{\rho}}=\mathbf{0}$; on the other hand, if $q_{f,l,m}=1$, $\ensuremath{\overline{\textrm{C4}}}$ is inactive due to C5. Moreover, for a given BS cooperation decision $q_{f,l,m}$, $\ensuremath{\overline{\textrm{C4}}}$ reduces to a convex quadratic inequality constraint.

Next, let $\mathbf{W}_{\boldsymbol{\rho}} \triangleq \mathbf{w}_{\boldsymbol{\rho}} \mathbf{w}_{\boldsymbol{\rho}}^{H} \succeq \mathbf{0}$ and $\mathbf{H}_{\boldsymbol{\rho}} \triangleq \mathbf{h}_{\boldsymbol{\rho}} \mathbf{h}_{\boldsymbol{\rho}}^{H}$. The QoS constraint C6 can be transformed into affine inequality constraints, 
\begin{alignat}{1}
\textrm{C6}\iff & \Gamma_{\boldsymbol{\rho}} \ge \kappa_{\boldsymbol{\rho}}^{\mathrm{req}}  \label{eq:qos-con-trans1}
 \\
\iff & \overline{\textrm{C6}}:\frac{1}{\kappa_{\boldsymbol{\rho}}^{\mathrm{req}}}\mathrm{tr}\left(\mathbf{W}_{\boldsymbol{\rho}}\mathbf{H}_{\boldsymbol{\rho}}\right)\ge\sigma^{2}
+\sum_{\boldsymbol{\rho}' \neq\boldsymbol{\rho}} \mathrm{tr} \left(\mathbf{W}_{\boldsymbol{\rho}'} \mathbf{H}_{\boldsymbol{\rho}} \right), \; \forall \boldsymbol{\rho}, \nonumber
\end{alignat}
where $\kappa_{\boldsymbol{\rho}}^{\mathrm{req}} \triangleq2^{R_{\boldsymbol{\rho}}^{\mathrm{req}}}-1$. C6 and $\overline{\textrm{C6}}$ are equivalent if and only if the following constraint holds 
\begin{equation}
\textrm{C10:} \; \mathbf{W}_{\boldsymbol{\rho}}\succeq\mathbf{0} \textrm{ and } \mathrm{rank}(\mathbf{W}_{\boldsymbol{\rho}})\le1.
\end{equation}

Finally, Proposition \ref{prop:C7} is applied to transform C7.
\vspace{-0.2cm}
\begin{prop}
\emph{\label{prop:C7}Assume that C10 holds. If $\varepsilon_{\textrm{e}}>0$, the secrecy constraint C7 can be equivalently transformed into a convex linear matrix inequality (LMI) as}
\emph{
\begin{alignat}{1}
\widetilde{\textrm{C7}}\textrm{: } &\mathbf{U}_{\textrm{e}}^{H}(\mathbf{W}_{\boldsymbol{\rho}}- \frac{\kappa_{\boldsymbol{\rho}}^{\mathrm{tol}} }{\sigma_{\textrm{e}}^{2}} \mathbf{V} ) \mathbf{U}_{\textrm{e}} \preceq  \\
& \mathrm{diag} \Big((\kappa_{\boldsymbol{\rho}}^{\mathrm{tol}} - \delta_{\textrm{e}})\mathbf{1}_{1\times N_{\mathrm{e}}}, \frac{\delta_{\textrm{e}}}{\varepsilon_{\textrm{e}}^2} \mathbf{1}_{1\times MN_{\mathrm{t}}} \Big), \;\delta_{\textrm{e}}\ge0, \nonumber
\end{alignat}
where $\mathbf{U}_{\textrm{e}}=\big[\widehat{\mathbf{G}},\mathbf{I}_{MN_{\mathrm{t}}}\big]$ and $\kappa_{\boldsymbol{\rho}}^{\mathrm{tol}}\triangleq \sigma_{\textrm{e}}^{2} (2^{R_{\mathrm{e},\boldsymbol{\rho}}^{\mathrm{tol}}}-1 )$. Moreover, if $\varepsilon_{\textrm{e}}=0$, ${\textrm{C7}}$ is equivalent to
\begin{equation}
\textrm{\ensuremath{\overline{\textrm{C7}}}:}\quad\mathbf{G}^{H}\mathbf{W}_{\boldsymbol{\rho}}\mathbf{G}\preceq \sigma_{\textrm{e}}^{2}\kappa_{\boldsymbol{\rho}}^{\mathrm{tol}} \mathbf{Z}_{\mathrm{e},\boldsymbol{\rho}},
\end{equation}
where $\mathbf{Z}_{\mathrm{e},\boldsymbol{\rho}}$ is defined after \eqref{eq:ER-rate}.
}
\end{prop}
\vspace{-0.2cm}
\begin{IEEEproof}
Please refer to Appendix \ref{append4}.
\end{IEEEproof}
Now, by applying the above transformations and relaxing the rank constraint $\mathrm{rank}(\mathbf{W}_{\boldsymbol{\rho}})\le1$, i.e., removing it from C10, we obtain the following convex semidefinite program (SDP), 
\vspace{-.2cm}
\begin{align}
\textrm{R1: }\;\minimize%_{\mathbf{W}_{\boldsymbol{\rho}},\mathbf{V},\delta_{\boldsymbol{\rho}},\delta_{\textrm{e}}}
\; & \mathrm{tr}\big(\sum\nolimits _{\boldsymbol{\rho}\in\mathcal{S}}\mathbf{W}_{\boldsymbol{\rho}}+\mathbf{V}\big)\\
{{\st \;}} & \textrm{\ensuremath{\overline{\textrm{C4}}}: }\mathrm{tr}\left(\boldsymbol{\Lambda}_{m}\mathbf{W}_{\boldsymbol{\rho}}\right)\le q_{f,l,m}P_{m}^{\max},\; \forall m, \forall \boldsymbol{\rho},\nonumber \\
 & \textrm{\ensuremath{\overline{\textrm{C5}}}: }\mathrm{tr}\big(\boldsymbol{\Lambda}_{m}\big(\sum_{\boldsymbol{\rho}\in\mathcal{S}}\mathbf{W}_{\boldsymbol{\rho}}+\mathbf{V}\big)\big)\le P_{m}^{\max},\; \forall m,\nonumber \\
 & \textrm{\ensuremath{\overline{\textrm{C6}}}, \ensuremath{\widetilde{\textrm{C7}}}/\ensuremath{\overline{\textrm{C7}}}, }\textrm{\ensuremath{\overline{\textrm{C10}}}: }\mathbf{W}_{\boldsymbol{\rho}}\succeq\mathbf{0},\mathbf{V}\succeq\mathbf{0},\; \forall \boldsymbol{\rho},\nonumber \\
{\revise{\var}} \; & {\revise{\mathbf{W}_{\boldsymbol{\rho}},\mathbf{V}, \delta_{\textrm{e}},}} \nonumber
\end{align}
where $\widetilde{\textrm{C7}}$ and $\overline{\textrm{C7}}$ are applied when $\varepsilon_{\textrm{e}}>0$ and $\varepsilon_{\textrm{e}}=0$, respectively. Generally, problem R1 achieves a lower bound on the optimal value of R0($\mathbf{D}_{\text{\mbox{II}},2}$). If the solution of R1 further satisfies $\mathrm{rank}(\mathbf{W}_{\boldsymbol{\rho}})\le1,\;\boldsymbol{\rho}\in\mathcal{S}$, then the optimal solution of problem R0($\mathbf{D}_{\text{\mbox{II}},2}$) is readily available based on $\mathbf{W}_{\boldsymbol{\rho}} = \mathbf{w}_{\boldsymbol{\rho}}\mathbf{w}_{\boldsymbol{\rho}}^H$, i.e., the relaxation is tight. For the problem at hand, however, the relaxation is always tight, which is established in the following theorem.

%\vspace{-0.2cm} 
\begin{thm}[Rank of Optimal $\mathbf{W}_{\boldsymbol{\rho}}$]%[Hidden Convexity of Problem R0($\mathbf{D}_{\text{\mbox{II}},2}$)]
\emph{\label{prop2}Problems R0($\mathbf{D}_{\text{\mbox{II}},2}$) and R1 are equivalent in the sense that both problems have the same optimal value; in particular, the optimal solution $\mathbf{W}_{\boldsymbol{\rho}}^{*}$ of R1 satisfies $\mathrm{rank}(\mathbf{W}_{\boldsymbol{\rho}}^{*})\le1,\;\boldsymbol{\rho}\in\mathcal{S}$, and the optimal beamforming vector $\mathbf{w}_{\boldsymbol{\rho}}^{*}$ of R0($\mathbf{D}_{\text{\mbox{II}},2}$) is given by the principal
eigenvector of $\mathbf{W}_{\boldsymbol{\rho}}^{*}$. This result holds for $\varepsilon_{\textrm{e}}\ge0$.}
\end{thm}
\vspace{-0.2cm}
\begin{IEEEproof}
Please refer to Appendix \ref{append2}. 
\end{IEEEproof}
Based on Theorem~\ref{prop2}, problem R0($\mathbf{D}_{\text{\mbox{II}},2}$) can be efficiently solved using standard convex optimization algorithms. For example, the interior-point method \cite{Bertsekas82Lagrange,Ye2011interior}, which is implemented in existing numerical convex program solvers such as CVX \cite{cvx}, is applicable. The resulting computational complexity of solving problem R0($\mathbf{D}_{\text{\mbox{II}},2}$) with respect to the number of LRs, $K$, the number of eavesdropping antennas, $N_{\mathrm{e}}$, and the total number of transmit antennas, $M N_{\mathrm{t}}$, is given by \cite[Theorem 3.12]{polik10interior}
\vspace{-.2cm}
\begin{equation}
\Theta^{\mathrm{sdp}} =\mathcal{O}\left( \underset{\textrm{Number of iterations}}{\underbrace{\sqrt{\Phi}\log\left({\epsilon}^{-1}\right)}} \; \underset{\textrm{Complexity per iteration}}{\underbrace{\left(\Xi\Phi^{3}+\Xi^{2}\Phi^{2}+\Xi^{3}\right)}} \right), 
\label{sdp-comp}
\end{equation}
with $\Phi \!\triangleq\! MN_{\mathrm{t}}(K\!+\!1)$ and $\Xi \!\triangleq\! MK\!+\!M\!+\!2K$. Herein, $\epsilon \!>\! 0$ is the desired solution accuracy specified for the adopted numerical solver and $\mathcal{O}(\cdot)$ denotes the big-$O$ notation. 
%The first term  represents the number of iterations required to achieve a solution with \epsilon-optimality The second term is the computational complexity required per iteration. 
Based on \eqref{sdp-comp}, the algorithm for solving problem R0($\mathbf{D}_{\text{\mbox{II}},2}$) has a polynomial-time computational complexity, which is desirable for real time implementation \cite{cvx}. 

\vspace{-0.4cm}
\subsection{\label{sub4-2}General Case: Greedy Iterative Solution of R0}

For limited cache capacity, problem R0 is NP-hard due to the non-convex constraints C2 and C3. The optimal solution has to be determined by enumerating all possible cooperation sets satisfying the backhaul capacity constraint C3. For this purpose, enumeration methods such as exhaustive search and branch-and-bound \cite{Floudas1995MINLP} are applicable. However, although the remaining cooperative beamforming problem, i.e., problem R0($\mathbf{D}_{\text{\mbox{II}},2}$), can be efficiently solved for each choice of cooperation sets, cf. Theorem \ref{prop2}, the overall computational complexity grows exponentially with the number of BSs due to the combinatorial nature of the problem. To be specific, we define $\overline{T}_{m}\triangleq\min\left\{ \left\lfloor B_{m}^{\max}/\min_{f\in\mathcal{F}(\mathcal{S})}Q_{f,m}\right\rfloor ,F(\mathcal{S})\right\} $ and $\underline{T}_{m}\triangleq\left\lfloor B_{m}^{\max}/\max_{f\in\mathcal{F}(\mathcal{S})}Q_{f,m}\right\rfloor $. According to Lemma \ref{lem:(Monotonicity-of-R0)}, the optimal cooperation formation solutions are contained in the vertices of the polyhedral simplex defined by the intersection of hyperplanes $\sum_{f\in\mathcal{F}(\mathcal{S})}q_{f,l,m}\le T_{m}$ and hypercubes $q_{f,l,m}\in[0,1]$, where $\underline{T}_{m}\le T_{m}\le\overline{T}_{m},m\in\mathcal{M}$. As a result, the enumeration of approximately $\prod_{m=1}^{M}\left(_{F(\mathcal{S})}^{\;T_{m}}\right)$ vertices %choices of the
(cooperation sets) is required in the worst case for solving problem R0 by exhaustive search (branch-and-bound).

Non-polynomial time enumeration methods are only applicable for systems of small size. For practical systems with medium and large sizes, however, effective polynomial-time algorithms are desired. %preferred. 
Herein, a low-complexity iterative algorithm based on greedy heuristics is proposed to solve problem R0 and is summarized in Algorithm \ref{alg1}.

Let $k$ be the iteration index. Define $\mathcal{Q}_{k}\triangleq\left\{ (f,m)\mid q_{f,l,m} =1, m\in\mathcal{M}, f\in F(\mathcal{S})\right\} $ as the BS cooperation solution set at iteration $k$. Algorithm \ref{alg1} starts with the initialization $\mathcal{Q}_{0}  \! =  \! \prod_{ f \in \mathcal{F}(\mathcal{S})}\mathcal{M}_{f,l}^{\mathrm{F-Coop}}$ $= \prod_{f\in\mathcal{F}(\mathcal{S})}\mathcal{M}$, that is, $q_{f,l,m} =1, \forall m\in\mathcal{M}, \forall f\in F(\mathcal{S})$.
At iteration $k=1,2,\ldots$, the $q_{f,l,m}$'s are fixed according to $\mathcal{Q}_{k-1}$ and, consequently, the %remaining 
cooperative beamforming solutions are obtained via problem R0($\mathbf{D}_{\text{\mbox{II}},2}$), whose optimal value is denoted by $f_{\mathrm{\text{\mbox{II}}}}^{*}(\mathcal{Q}_{k-1})$. If $\mathcal{Q}_{k-1}$ fulfills the backhaul capacity constraint C3, the algorithm stops and returns the solutions for cooperative BS transmission. Otherwise, the greedy algorithm sets $q_{f',l,m'}=0$ for that $(f',m')\in\mathcal{Q}_{k-1}$ which incurs the smallest penalty on the total transmit power, i.e., 
\vspace{-.2cm}
\begin{equation}
(f',m') \in \!\! \underset{(f,m)\in\mathcal{F}(\mathcal{S})\times\mathcal{M}_{k}^{\mathrm{vio}}}{\arg\min} \!\!
\left[f_{\mathrm{\text{\mbox{II}}}}^{*}(\mathcal{Q}_{k-1}\backslash\left\{ (f,m)\right\} )-f_{\mathrm{\text{\mbox{II}}}}^{*}(\mathcal{Q}_{k-1})\right],\label{eq:greedy}
\end{equation}
and updates $\mathcal{Q}_{k} =\mathcal{Q}_{k-1}\backslash\left\{ (f',m')\right\}$. In \eqref{eq:greedy}, $\mathcal{M}_{k-1}^{\mathrm{vio}}$ denotes the index set of BSs violating constraint C3 if $\mathcal{Q}_{k-1}$ is adopted for cooperative transmission, i.e., 
\vspace{-.2cm}
\begin{equation}
\mathcal{M}_{k-1}^{\mathrm{vio}}\triangleq \Big\{ m\in\mathcal{M}\mid\sum_{(f,m)\in\mathcal{Q}_{k-1}}Q_{f,m}>B_{m}^{\max}\Big\} .\label{eq:vioBS}
\end{equation}
The iteration process is repeated until C3 is fulfilled.

Note that during each iteration of Algorithm \ref{alg1}, \eqref{eq:greedy} is solved by enumerating $F(\mathcal{S})\times\left|\mathcal{M}_{k-1}^{\mathrm{vio}}\right|$ choices of $(f,m)$. The total number of choices is bounded from above by $F(\mathcal{S})\times\left|\mathcal{M}_{0}^{\mathrm{vio}}\right|\times T$ in the worst case, where $\sum_{m\in\mathcal{M}}\underline{T}_{m}\le T\le\sum_{m\in\mathcal{M}}\overline{T}_{m}$. Consequently, the overall computational complexity of Algorithm \ref{alg1} is given by
\vspace{-.2cm}
\begin{equation}
\Theta^{\mathrm{greedy}} = F(\mathcal{S})\times\left|\mathcal{M}_{0}^{\mathrm{vio}}\right|\times T  \Theta^{\mathrm{sdp}},
\end{equation}
which grows only polynomially with the number of BSs $M$ and the number of LRs $K$, cf. \eqref{sdp-comp}. In general, the proposed greedy algorithm is suboptimal. However, for large cache capacity, Algorithm \ref{alg1} terminates without having to solve \eqref{eq:greedy} and the obtained solution is globally optimal.

\begin{algorithm}[t]
\protect\protect\caption{\textcolor{black}{Greedy Iterative Algorithm for Solving R0} }

\label{alg1} \small \begin{algorithmic}[1] 

\STATE \textbf{Initialization}: $\mathcal{Q}_{0}\leftarrow\prod_{f\in\mathcal{F}(\mathcal{S})}\mathcal{M}_{f,l}^{\mathrm{F-Coop}}$
, $k\leftarrow1$;

\STATE Solve problem R0($\mathbf{D}_{\text{\mbox{II}},2}$) for $\mathcal{Q}_{0}$;

\WHILE{$\mathcal{M}_{k-1}^{\mathrm{vio}}\neq\emptyset$ (cf. (\ref{eq:vioBS}))}

\FOR{\textbf{each} $(f,m)\in\mathcal{F}(\mathcal{S})\times\mathcal{M}_{k}^{\mathrm{vio}}$}

\STATE Solve problem R0($\mathbf{D}_{\text{\mbox{II}},2}$) for $\mathcal{Q}_{k-1}\backslash\left\{ (f,m)\right\} $;

\ENDFOR

\STATE $\mathcal{Q}_{k}\leftarrow\mathcal{Q}_{k-1}\backslash\left\{ (f',m')\right\} $,
where $(f',m')$ solves (\ref{eq:greedy});

\STATE $k\leftarrow k+1$, 

\ENDWHILE 

\end{algorithmic} 
\end{algorithm}

\begin{rem}
We note that alternative polynomial-time methods that could be used to solve problem R0 have certain pitfalls. For example, by reformulating the binary variables as $\ell_{0}$-norms, problem R0 can be solved by the approximation methods proposed in \cite{Zhao13TWC:BchOpt,Zhuang14TSP:AsyCoop}. However, the solutions obtained by these approximation methods are generally (primal) infeasible for most of the cases under investigation. This is because, on the one hand, the big-M constraint $\ensuremath{\overline{\textrm{C4}}}$ would force $q_{f,l,m}$ to small non-zero fractions instead of binary solutions that fulfill C2. On the other hand, generating primal feasible solutions from approximate results is a non-trivial task for NP-hard problems. For example, a brute-force deterministic or random rounding of these continuous solutions usually leads to the violation of the backhaul capacity constraint C3. {\revise{For similar reasons, primal feasibility is also not ensured when other polynomial-time methods such as convex relaxation (e.g., linear programming or SDP based relaxation of the binary constraints) and difference of convex programming (also referred to as successive convex approximation) \cite{ScutariTSP15:SCP} are adopted.}} Yet, with Algorithm~\ref{alg1}, the likelihood of obtaining an infeasible primal solution is low due to the greedy iterative search, cf. \eqref{eq:vioBS}. As will be shown in Section \ref{sec:Simulation-Results}, the solution obtained with Algorithm~\ref{alg1} is close to the optimal value in the medium and large cache capacity regimes. 
\end{rem}

\vspace{-0.2cm}
\subsection{Solution of Problem Q0\label{sec4-3}}

Problem Q0 has a considerably ($\Omega$-times) larger problem size than R0. Hence, solving problem Q0 via enumeration methods seems impossible due to the overwhelming computational complexity. Besides, the greedy suboptimal method  in Algorithm~\ref{alg1} cannot be directly applied for solving problem Q0 since constraint C1 becomes bilinear over the joint optimization space $\left\{ \mathbf{C}_{\text{\mbox{I}}}\right\}$ of Q0. We address both issues by applying the following binary relaxation method.

In particular, the bilinear constraint C1 is transformed to 
\begin{equation}
\widetilde{\textrm{C1}}: c_{f,m}+b_{f,l,m,\omega}\ge q_{f,l,m,\omega}.
\end{equation}
If the average backhaul capacity is insufficient, $\widetilde{\textrm{C1}}$ and C8 together lead to $b_{f,l,m,\omega}=(1-c_{f,m})\times q_{f,l,m,\omega}$ since Q0 enjoys a similar monotonicity as R0, cf. Lemma \ref{lem:(Monotonicity-of-R0)}; otherwise, $\widetilde{\textrm{C1}}$ is inactive. Thus, $\widetilde{\textrm{C1}}$
and C1 are equivalent.

Moreover, let $\widetilde{\textrm{C2}}$ be a relaxation of C2 where the binary constraints are replaced by $q_{f,l,m,\omega}\in[0,1]$. By adopting $\widetilde{\textrm{C1}}$ and $\widetilde{\textrm{C2}}$, we arrive at a relaxed version of Q0:
\begin{align}
\textrm{Q1: }\minimize%_{\mathbf{C}_{\text{\mbox{I}}}}
\;\; & \frac{1}{\Omega}\sum\nolimits _{\omega=1}^{\Omega}\; f_{\mathrm{\text{\mbox{I}}},\omega}\\
{\st}\;\; & \ensuremath{\overline{\textrm{C3}}}, \textrm{C8}, \;\;
\overline{\textrm{C9}}{\textrm{: }}
\mathbf{D}_{\text{\mbox{I}},\omega}\in\widehat{\mathcal{D}}_{\text{\mbox{I}},\omega},\;\omega\in\left\{ 1,\ldots,\Omega\right\} ,\nonumber \\
{\revise{\var}} \;\; & {\revise{ \mathbf{C}_{\text{\mbox{I}}} = [c_{f,m},\,q_{f,l,m,\omega},b_{f,l,m,\omega},\mathbf{w}_{\boldsymbol{\rho},\omega},\mathbf{V}_{\omega}], }} \nonumber
\end{align}
where $\widehat{\mathcal{D}}_{\text{\mbox{I}},\omega}\triangleq\left\{ \mathbf{D}_{\text{\mbox{I}},\omega}\mid \widetilde{\textrm{C1}},  \widetilde{\textrm{C2}} \textrm{, C4--C7}\right\} $. Although problem Q1 remains non-convex due to constraints C4, C6, and C7, %the hidden convexity of problem Q1 can be shown in a similar manner as that of R0 in Theorem \ref{prop2}.
an equivalent convex SDP can be obtained for Q1 in a similar manner as for problem R0, cf. Section \ref{sub4-1}. Thus, the relaxed problem Q1 can be solved efficiently (in polynomial time) using the interior point method~\cite{Bertsekas82Lagrange}. 

In general, the proposed relaxation solution of Q1 provides a performance lower bound for Q0. However, the following theorem establishes that the relaxation solution is asymptotically optimal in the limiting case of   large $\Omega$.
%\vspace{-0.2cm}
\begin{thm}[Asymptotic Optimality of the Relaxation Solution]
\label{thm2} \emph{Problems Q1 and Q0 become equivalent as $\Omega\to\infty$ in the sense that their optimum values and the optimal caching decisions become identical if problem Q0 is feasible.} 
\end{thm}
\vspace{-0.2cm}
\begin{IEEEproof}
Please refer to Appendix~\ref{append}. 
\end{IEEEproof}

\vspace{-0.4cm}
\section{\label{sec:Simulation-Results}Simulation Results}

In this section, we evaluate the system performance for the proposed caching and secure delivery schemes. Consider a cluster of \textbf{$M=7$} hexagonal cells, where a BS is deployed at the center of each cell with an inter-BS distance of $500$~m. Each BS is equipped with $N_{\mathrm{t}}=4$ antennas and the ER has $N_{\mathrm{e}}=2$ antennas. We assume that a library of $F = 10$ video files, each of duration $45$ minutes and size $500$ MB (Bytes), is delivered to $K=5$ single-antenna LRs. Consequently, an estimated secrecy data rate of $R_{\boldsymbol{\rho}}^{\mathrm{sec}} \approx Q_{f} = 500 \times 8.0 \times 10^{6}/(45 \times 60) \approx1.5$~Mbps is required at each LR for uninterrupted video streaming. The LRs and the ER are uniformly and randomly distributed in the system while the minimum distance between receiver and BS is $50$~m. Each LR requests one file independent of the other LRs. Let $\theta_{f}$ be the probability of file $f\in\mathcal{F}$ being requested and let $\boldsymbol{\theta}=[\theta_{1},\ldots,\theta_{F}]$ be the probability distribution of the requests for the different files. We set $\theta_{f}=\frac{1}{f^{\kappa}}/\sum_{f \in \mathcal{F}}\frac{1}{f^{\kappa}}$ with $\kappa=1.1$ according to the Zipf distribution \cite{Breslau99Zipf}. Moreover, the 3GPP path loss model (``Urban Macro NLOS'' scenario) in \cite{3GPP:TR36814} is adopted. For channel estimation, we define the normalized estimation error of the ER as $\alpha^2 \triangleq \tfrac{\varepsilon_{\textrm{e}}^2}{\left\Vert \mathbf{G}\right\Vert _{F}^2} $. Unless otherwise specified, we assume $\alpha^2 = 0.05$. The capacities of the backhaul links are independently and identically distributed (i.i.d.) as $\mathrm{Pr}(B_{m}^{\max}=0\textrm{ Mbps})=0.3$, $\mathrm{Pr}(B_{m}^{\max}=3\textrm{ Mbps})=0.4$, and $\mathrm{Pr}(B_{m}^{\max} = 6 \textrm{ Mbps})=0.3$, $\forall m$, which can be interpreted as the probabilities of high, medium, and low non-VoD traffic scenarios in the cellular network, respectively. The other relevant system parameters are given in Table~\ref{tab1}. Before video delivery starts, $\Omega=50$ scenarios are randomly generated based on the adopted models for the user preference, CSI, and backhaul capacity to determine the initial cache status, cf. problem~Q0. 

\begin{table}
\centering\protect\protect\caption{\label{tab1}Simulation parameters.}
\vspace{-.3cm}
\small 
\begin{tabular}{|c|c|}
\hline 
 Parameters  & Settings \tabularnewline
\hline 
\hline 
{System bandwidth } & {10 MHz}\tabularnewline
\hline 
{Duration of time slot } & {$\tau=$ 10 ms}\tabularnewline
\hline 
{File splitting } & {$L=45\textrm{ mins}/\tau=2.7\times10^{4}$ }\tabularnewline
\hline 
{BS transmit power } & {$P_{m}^{\max}=$ 48 dBm}\tabularnewline
\hline 
{Noise power density } & {$-$172.6 dBm/Hz}\tabularnewline
\hline 
{Delivery QoS requirement } & {$R_{\boldsymbol{\rho}}^{\textrm{req}}=1.1\times R_{\boldsymbol{\rho}}^{\mathrm{sec}}=$
1.65 Mbps }\tabularnewline
\hline 
{Delivery secrecy threshold } & {$R_{\mathrm{e},\boldsymbol{\rho}}^{\mathrm{tol}}=0.1\times R_{\boldsymbol{\rho}}^{\textrm{sec}}=$
0.15 Mbps}\tabularnewline
\hline 
\end{tabular}
\end{table}

\vspace{-0.4cm}
\subsection{Performance of the Proposed Caching Scheme}
First, we study the performance of the proposed caching scheme. For comparison, we consider three heuristic caching schemes as baselines:
\begin{itemize}
\item Baseline 1 (Preference-based caching): The most popular files are cached. Assuming $\boldsymbol{\theta}$ is known, the cache control decision is made based on 
\vspace{-.2cm}
\begin{align}\maximize%_{c_{f,m}} 
& \quad\sum\nolimits_{f\in\mathcal{F}, \, m\in\mathcal{M}}{\theta_{f} c_{f,m} V_{f}}\\
{\st} & \quad c_{f,m}\in[0,1], f\in \mathcal{F}, m\in \mathcal{M}, \;\; \textrm{C8}, \nonumber \\
{\revise{\var}}  & \quad {\revise{c_{f,m}.}} \nonumber
\end{align}

\item Baseline 2 (Uniform caching): The same amount of data is cached for each file, i.e., 
\begin{equation}
\vspace{-.1cm}
c_{f,m}V_{f}=\frac{1}{F} \min\{C_{m}^{\max},\,\sum_{f'=1}^{F}V_{f'}\},\; \forall f, \forall m,
\end{equation}
and the users' preferences are not taken into account. 

{\revise{\item Baseline 3 (Power-efficient caching): This scheme is identical to the proposed caching scheme, except that the secrecy constraint C7 in Q0 is excluded from $\mathcal{D}_{\mathrm{I},\omega}$ during cache placement. }}
\end{itemize}
{\revise{For Baselines 1, 2, and 3, the proposed delivery scheme, i.e., Algorithm~\ref{alg1}, is adopted.}}

\begin{figure*}[t]
\vspace{-0.4cm}
 \centering 
\subfloat[]{\includegraphics[width=3.2in]{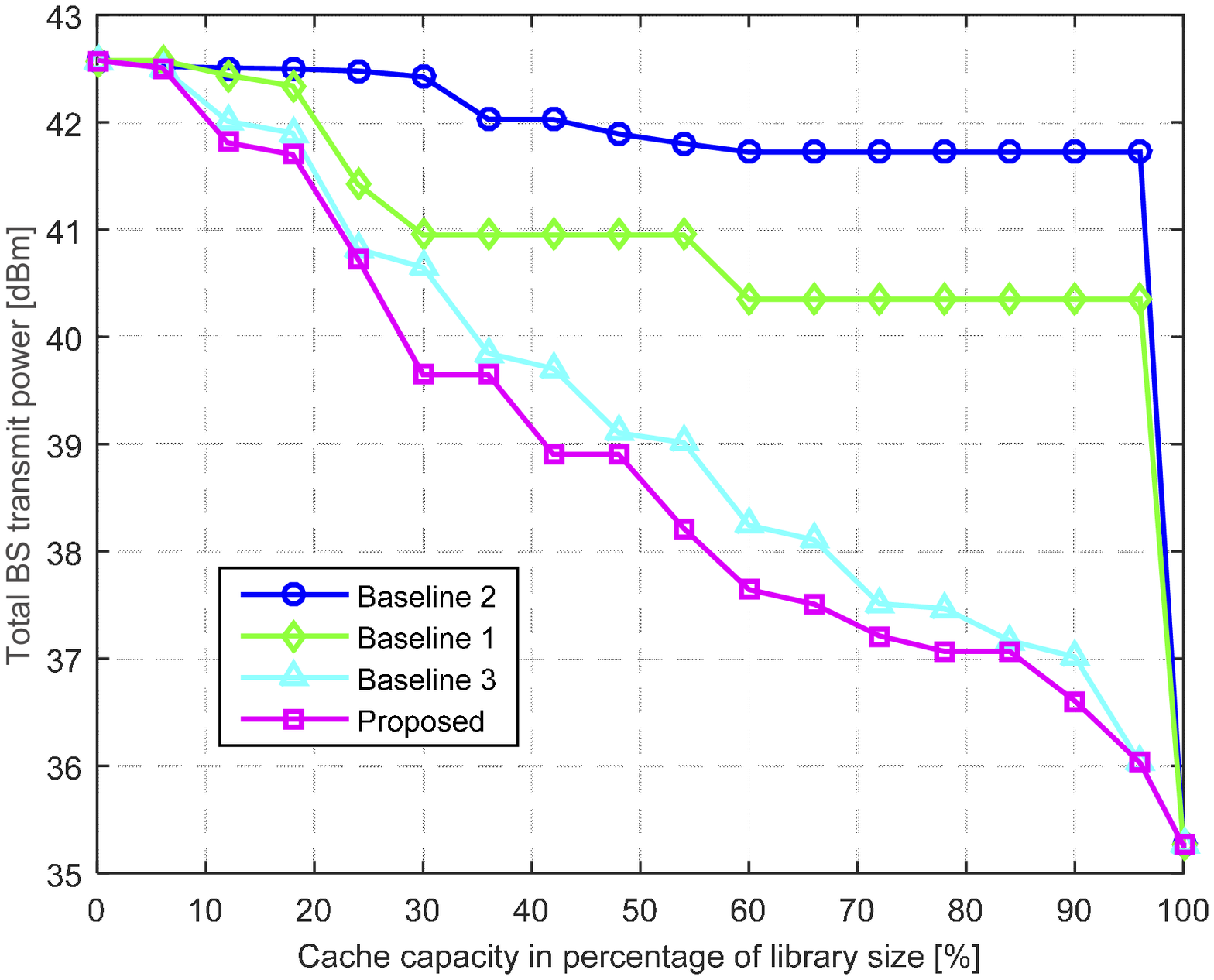} \label{fig:cach1}} 
\subfloat[]{\includegraphics[width=3.2in]{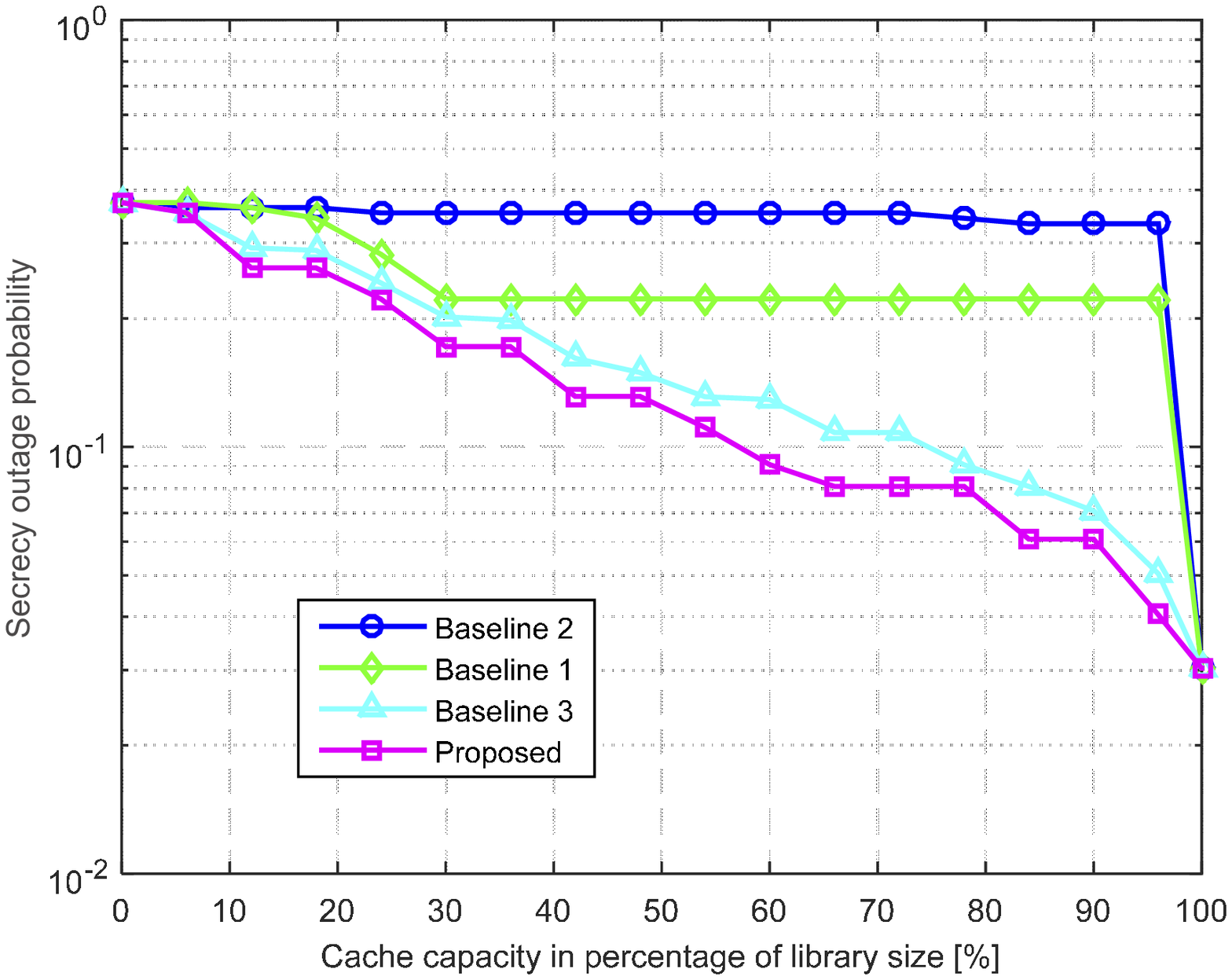} \label{fig:cach2}}
\vspace{-0.2cm}
\caption{\small {\revise{(a) Total BS transmit power and (b) secrecy outage probability versus cache capacity (in percentage of the total library size) for different caching schemes.}} }
\end{figure*}

Figs.~\ref{fig:cach1} and \ref{fig:cach2} show the total transmit power and the secrecy outage probability, defined as $p_{\textrm{out}}\triangleq\mathrm{Pr}(R_{\boldsymbol{\rho}}^{\mathrm{sec}} < [R_{\boldsymbol{\rho}}^{\textrm{req}}-R_{\mathrm{e},\boldsymbol{\rho}}^{\mathrm{tol}}]^{+})$, of the considered caching schemes as functions of the cache capacity, respectively.  Herein, $p_{\textrm{out}}$ corresponds to the probability that problem R0 is infeasible because either the scheme fails to satisfy the QoS constraint C6 or the secrecy constraint C7. As can be observed from Figs.~\ref{fig:cach1} and \ref{fig:cach2}, a larger cache capacity leads to both a lower total BS transmit power and a smaller secrecy outage probability\footnote{In case of outage, in a practical system, a retransmission request would be triggered.} as larger (virtual) transmit antenna arrays can be formed during video delivery. For example, for the proposed scheme, the average number of cooperating BSs is $4.0$ for $C_{m}^{\max}=4000$ MB ($80\%$ of library size) compared to $2.8$ for $C_{m}^{\max}=1000$ MB ($20\%$ of library size), which leads to a transmit power reduction of up to $4.8$ dB. The performance gap between the considered caching schemes is negligible for small (large) cache capacities because of insufficient (saturated) BS cooperation. For medium cache capacities, however, the proposed caching scheme achieves considerable transmit power savings due to its ability to exploit the historical information regarding user requests, backhaul capacity, and CSI for resource allocation. {\revise{For a similar reason, both the transmit power and the secrecy outage probability of Baseline 3 are lower than those of Baselines 1 and 2. However, the performance of Baseline 3 is generally worse than that of the proposed scheme. This indicates that including the secrecy constraint also for cache placement is necessary for maximizing the performance during video file delivery.}}
Note also that when the increase in cache capacity is not sufficient to support additional BSs for cooperative transmission, the performance does not improve. As a result, the total transmit power in Fig.~\ref{fig:cach1} decreases in a piece-wise constant manner as the cache capacity increases.

\vspace{-0.4cm}
\subsection{Performance of the Proposed Delivery Scheme}
Next, we study the performance of the proposed delivery scheme. As a performance benchmark, the optimal solution of problem R0 is evaluated by an exhaustive search. Besides, the following naive delivery schemes are considered as baselines: 
\begin{itemize}
\item Baseline 4 (Coordinated beamforming): The user is associated with the nearest BS which has sufficient backhaul capacity available. Each video (sub)file is only delivered from the associated BS, i.e., $\sum_{m\in\mathcal{M}}q_{f,l,m}=1,\forall(f,l) \in \mathcal{F} \times \mathcal{L}$.

\item Baseline 5 (Full BS cooperation): The backhaul capacity constraints are dropped and all BSs cooperate to serve all users, i.e., $q_{f,l,m}=1,\forall f,l,m$. For Baselines 4 and 5, the optimal beamforming solutions are obtained based on R0($\mathbf{D}_{\text{\mbox{II}},2}$) where $q_{f,l,m}$s are fixed accordingly.

\item Baseline 6 (Non-robust transmission): Different from the proposed delivery scheme, the CP treats the possibly erroneous channel estimate $ \widehat{\mathbf{G}}$ as accurate. Consequently, the delivery decisions are made according to R0 by setting $\varepsilon_{\textrm{e}}^2 = 0$, irrespective of the channel estimation errors.
 
\end{itemize}

\begin{figure*}[t]
\vspace{-0.4cm}
 \centering 
\subfloat[]{\includegraphics[width=3.2in]{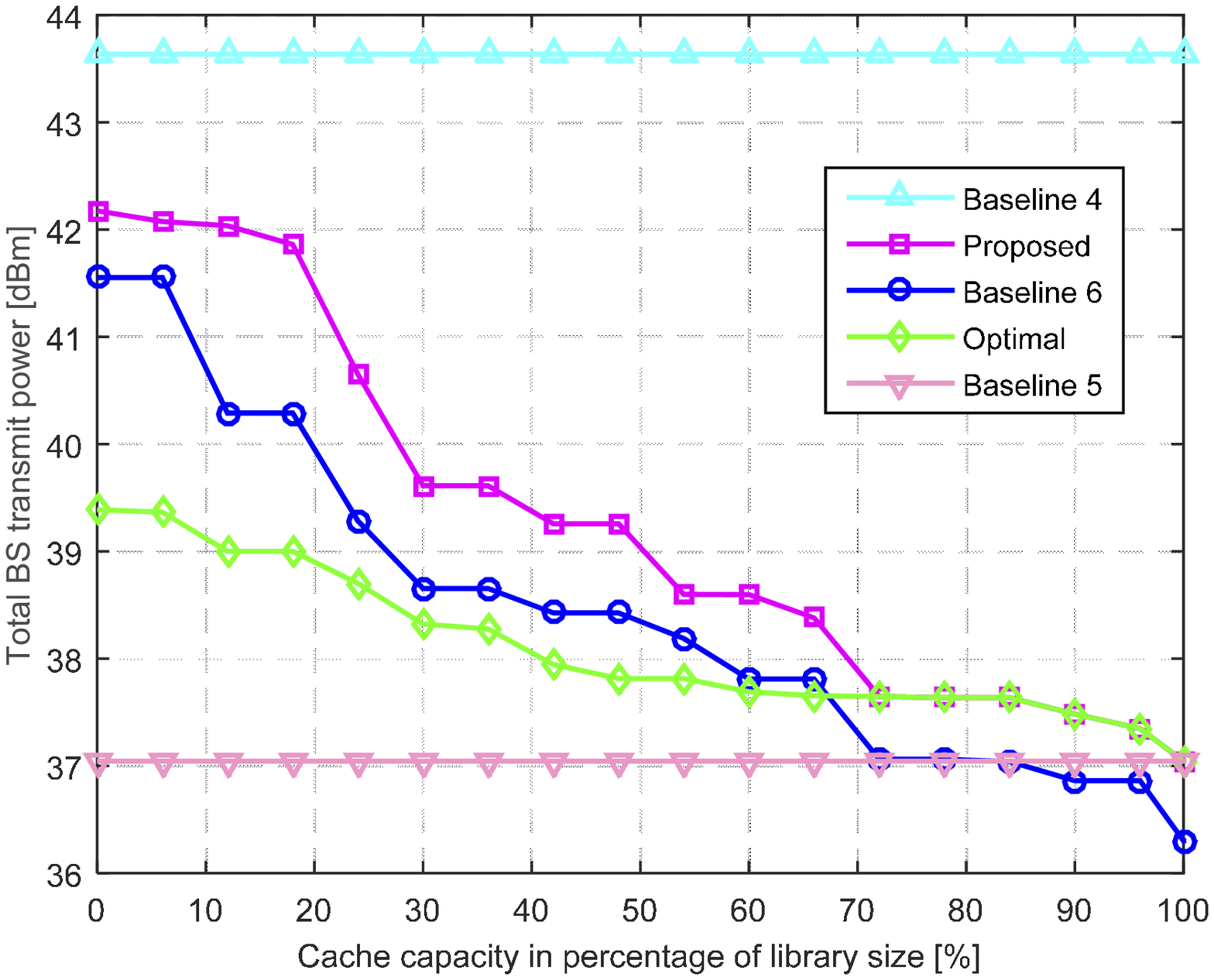} \label{fig:del1}}  
\subfloat[]{\includegraphics[width=3.2in]{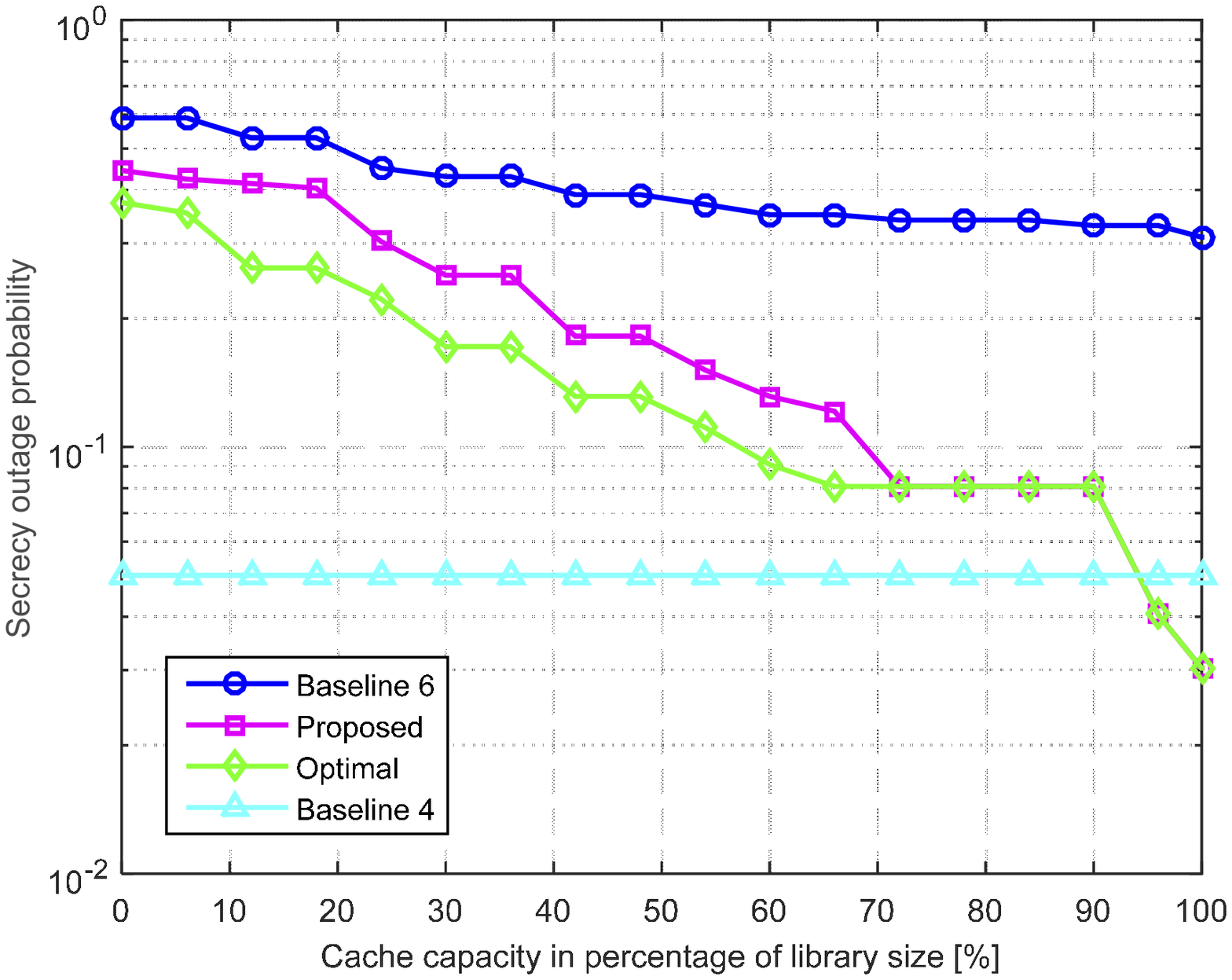} \label{fig:del2}}
\vspace{-0.2cm}
\protect\protect\caption{\small {\revise{(a) Total BS transmit power and (b) secrecy outage probability versus cache capacity (in percentage of the total library size) for different delivery schemes.}}}
\end{figure*}

Figs.~\ref{fig:del1} and \ref{fig:del2} illustrate the performance of the considered delivery schemes as functions of the cache capacity\footnote{The secrecy outage probability for Baseline 5 is not shown as Baseline 5 is infeasible for most cache capacity values.}. The initial cache status is determined by solving problem Q0. From Fig.~\ref{fig:del1}, we observe that, as expected, Baselines 4 and 5 constitute performance lower and upper bounds for the proposed delivery scheme, respectively. Comparing the proposed delivery scheme and the optimal delivery scheme, the performance gap between them reduces as the cache capacity increases. This is because, for large cache capacities, less backhaul traffic is generated and correspondingly the probability that C3 is active is reduced. It is interesting to observe that for a cache capacity of 3600~MB (70\% of library size), the proposed scheme already achieves the same performance as the optimal delivery scheme. 
Meanwhile, Fig.~\ref{fig:del1} shows that the total transmit power of Baseline 6 is lower than that of the proposed scheme. In the large cache capacity regime, Baseline 6 even consumes less transmit power than full BS cooperation. This is because, under imperfect CSI, cooperative beamforming alone is not sufficient to prevent data leakage to the ER. Hence, the proposed scheme and Baseline 5 have to transmit a non-negligible amount of AN to degrade the reception of the ER for enhancing communication secrecy. On the other hand, as the AN can also leak into the LR channels, cf. \eqref{eq:LR-rate}, a higher transmit power is also needed for cooperative beamforming to fulfill the QoS constraint C6. Nevertheless, we observe from Fig.~\ref{fig:del2} that, by consuming extra transmit power under imperfect CSI, the secrecy outage probability of the proposed and the optimal schemes can be kept low and decreases significantly with the cache capacity. In contrast, for Baseline 6, the BSs transmit with insufficient power since the imperfect CSI is treated as perfect CSI. %For example, the AN is generally not transmitted by Baseline 5. However, 
As a consequence, the secrecy outage probability for Baseline 6 is the highest among all the considered schemes and decreases much slower for increasing cache capacity than that of the proposed scheme. This implies that the imperfection of CSI has to be carefully taken into account for guaranteeing the secrecy of video delivery.  

\begin{figure}[t]
%\vspace{-0.4cm}
 \centering 
\subfloat[]{\includegraphics[width=3.2in]{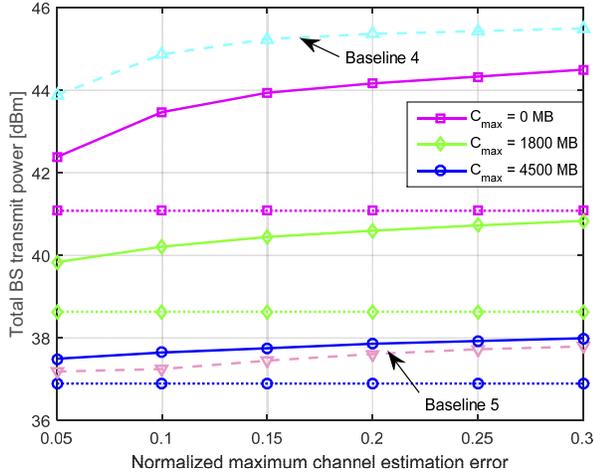} \label{fig:est1}}  \quad
\subfloat[]{\includegraphics[width=3.2in]{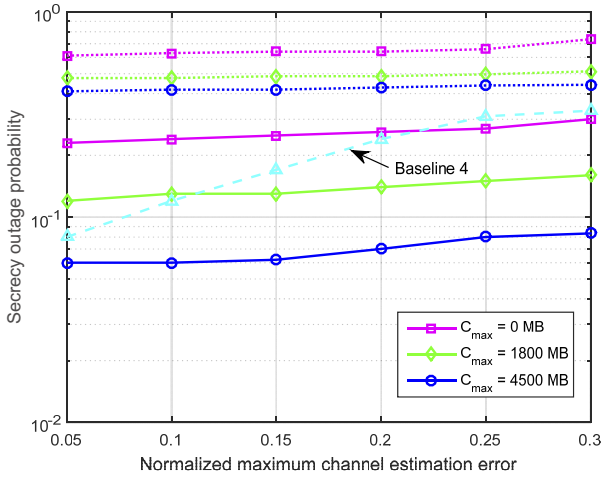} \label{fig:est2}}
\vspace{-0.2cm}
\caption{\small (a) Total BS transmit power and (b) secrecy outage probability of different delivery schemes versus normalized channel estimation error: the performances of the proposed scheme and Baseline 6 are shown as solid and dotted lines, respectively; the performances of Baselines 4 and 5 are as dashed lines, respectively. }
\end{figure}

\vspace{-0.4cm}
\subsection{Impact of Channel Estimation Errors}
In Figs.~\ref{fig:est1} and \ref{fig:est2}, we show the performance of the considered delivery scheme versus the normalized channel estimation error $\alpha^2$ for different cache capacities. As can be observed, for the proposed scheme, Baseline 4, and Baseline 5, both the total transmit power and the secrecy outage probability increase with the value of $\alpha^2$. This is because, as the uncertainty region $\mathcal{U}_{\mathrm{e}}$ is enlarged, the secrecy constraint C7 becomes more stringent, and to satisfy C7, a higher AN power is needed. Meanwhile, to satisfy the QoS constraint C6, a higher beamforming power is also needed to combat the leaked AN at the LRs. Nevertheless, by consuming extra transmit power, the proposed delivery scheme is able to maintain a reasonably low secrecy outage probability despite the deteriorating CSI quality. Moreover, for large cache capacities, the increase in transmit power is significantly reduced because a large cooperative transmit antenna array can be formed, and the performance gap between the proposed scheme and Baseline 5 becomes negligible. On the other hand, as Baseline 4 has only limited spatial degrees of freedom, its secrecy outage probability increases significantly with $\alpha^2$ even though a large transmit power is consumed. As for Baseline 6, although the transmit power remains constant as $\alpha^2$ increases, a high secrecy outage probability results, cf. Fig. \ref{fig:est2}.

\begin{figure}[t]
%\vspace{-0.4cm}
 \centering 
\subfloat[]{\includegraphics[width=3.2in]{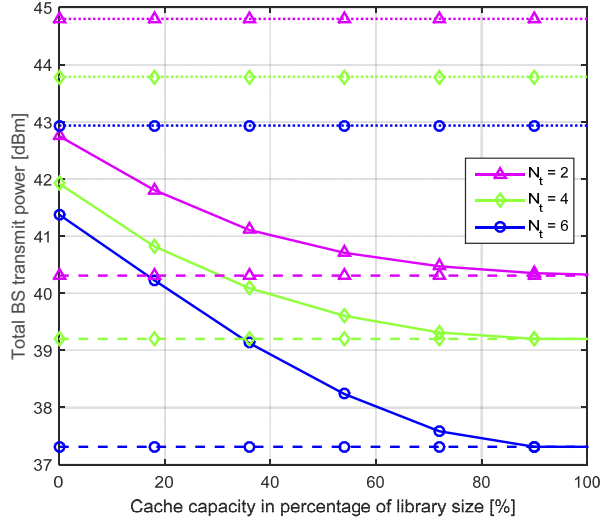}  \label{fig:ant1}}  \quad
\subfloat[]{\includegraphics[width=3.2in]{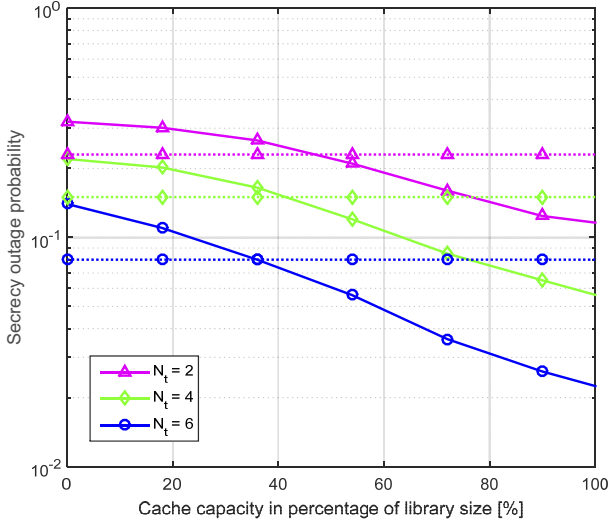} \label{fig:ant2}}
\vspace{-0.2cm}
\caption{\small{\revise{(a) Total BS transmit power and (b) secrecy outage probability of the proposed delivery scheme (solid line), Baseline 4 (dotted line) and Baseline 5 (dashed line) versus cache capacity (in percentage of the total library size) for different numbers of transmit antennas at BSs.}}}
\end{figure}

\vspace{-0.4cm}
\subsection{Impact of the Number of Transmit Antennas}
In Figs.~\ref{fig:ant1} and \ref{fig:ant2}, the performance of the proposed delivery scheme, Baseline 4, and Baseline 5 versus the cache capacity is evaluated for different numbers of transmit antennas, respectively. For a given cache capacity, more transmit antennas $N_{{\rm {t}}}$ equipped at the BSs, which increase the s.d.o.f., lead to transmit power savings for the considered schemes in Fig.~\ref{fig:ant1}. For similar reasons, the secrecy outage probability of the considered schemes is also decreased in Fig. \ref{fig:ant2} by using larger $N_{{\rm {t}}}$. However, different from Baselines 4 and 5, a further reduction in both the secrecy outage probability and the total BS transmit power can be achieved for the proposed scheme by increasing the cache capacity at the BSs. Particularly, when the cache capacity is sufficiently large, the proposed scheme  achieves the same power efficiency as Baseline 5, while simultaneously yielding a small secrecy outage probability.

\section{\label{sec:Conclusion}Conclusion}

In this paper, caching was exploited to improve PLS for cellular video streaming. Caching and cooperative transmission were optimized based on a mixed-integer two-stage robust optimization problem for minimization of the total transmit power needed to secure video streaming under imperfect CSI knowledge. Caching was shown to reduce the backhaul capacity required for cooperative transmission among a large group of BSs and to increase the available s.d.o.f. 
 As the problem was shown to be NP-hard, suboptimal polynomial-time algorithms were developed to solve the problem efficiently. The optimality of the proposed algorithms was verified in the regimes of a large cache capacity and a large number of training scenarios, respectively. Simulation results showed that the proposed caching and delivery schemes can significantly enhance both the PLS and power efficiency of cellular video streaming. Moreover, by deploying caches with large capacities, both the performance advantages and the robustness of full BS cooperation can be reaped even for limited backhaul capacities and imperfect CSI knowledge.

%\newpage 
\appendices{}
\section{\label{append4}Proof of Proposition \ref{prop:C7}}
Constraint C7 is first transformed to an LMI as follows,
\begin{alignat}{1}
\textrm{C7}\stackrel{\textrm{(a)}}{\iff} & \; 1+\frac{1}{\sigma_{\textrm{e}}^{2}}\mathbf{w}_{\boldsymbol{\rho}}^{H}\mathbf{G}\mathbf{Z}_{\mathrm{e},\boldsymbol{\rho}}^{-1}\mathbf{G}^{H}\mathbf{w}_{\boldsymbol{\rho}}\le2^{R_{\mathrm{e},\boldsymbol{\rho}}^{\mathrm{tol}}}, \; \forall\Delta\mathbf{G}\in\mathcal{U}_{\textrm{e}},\nonumber \\
\stackrel{\textrm{(b)}}{\iff} & \; \mathrm{tr}\left(\mathbf{Z}_{\mathrm{e},\boldsymbol{\rho}}^{-\frac{1}{2}}\mathbf{G}^{H} \mathbf{w}_{\boldsymbol{\rho}} \mathbf{w}_{\boldsymbol{\rho}}^{H} \mathbf{G} \mathbf{Z}_{\mathrm{e},\boldsymbol{\rho}}^{-\frac{1}{2}}\right) \le \kappa_{\boldsymbol{\rho}}^{\mathrm{tol}} \nonumber \\
\stackrel{\textrm{(c)}}{\implies} &  \lambda_{\max}\left(\mathbf{Z}_{\mathrm{e},\boldsymbol{\rho}}^{-\frac{1}{2}}\mathbf{G}^{H}\mathbf{W}_{\boldsymbol{\rho}}\mathbf{G}\mathbf{Z}_{\mathrm{e},\boldsymbol{\rho}}^{-\frac{1}{2}}\right) \le \kappa_{\boldsymbol{\rho}}^{\mathrm{tol}} \label{eq:31}\\
\iff & \;\mathbf{Z}_{\mathrm{e},\boldsymbol{\rho}}^{-\frac{1}{2}}\mathbf{G}^{H} \mathbf{W}_{\boldsymbol{\rho}}\mathbf{G}\mathbf{Z}_{\mathrm{e},\boldsymbol{\rho}}^{-\frac{1}{2}} \preceq \kappa_{\boldsymbol{\rho}}^{\mathrm{tol}}\mathbf{I} \quad \nonumber \\
\iff & \textrm{\ensuremath{\overline{\textrm{C7}}}: }  \mathbf{G}^{H}\mathbf{W}_{\boldsymbol{\rho}}\mathbf{G} \preceq \kappa_{\boldsymbol{\rho}}^{\mathrm{tol}}\mathbf{Z}_{\mathrm{e},\boldsymbol{\rho}},\;\forall\Delta\mathbf{G}\in\mathcal{U}_{\textrm{e}},\nonumber 
\end{alignat}
where (a) follows from Sylvester's determinant identity $\det(\mathbf{I}+\mathbf{A}\mathbf{B})=\det(\mathbf{I}+\mathbf{B}\mathbf{A})$, (b) follows from the trace identity $\mathrm{tr}(\mathbf{A}\mathbf{B})=\mathrm{tr}(\mathbf{B}\mathbf{A})$, and (c) is due to the inequality $\lambda_{\max}(\mathbf{A})\le\mathrm{tr}(\mathbf{A})$ for $\mathbf{A} \succeq \mathbf{0}$ \cite{Horn2012Matrix} and the variable substitution $\mathbf{W}_{\boldsymbol{\rho}} = \mathbf{w}_{\boldsymbol{\rho}}\mathbf{w}_{\boldsymbol{\rho}}^{H}$. For (c), $\lambda_{\max}(\mathbf{A})=\mathrm{tr}(\mathbf{A})$ holds if $\mathrm{rank}(\mathbf{A})\le1$. Considering the variable transformation $\mathbf{W}_{\boldsymbol{\rho}}=\mathbf{w}_{\boldsymbol{\rho}}\mathbf{w}_{\boldsymbol{\rho}}^{H}$, we conclude that C7 and $\overline{\textrm{C7}}$ are equivalent if $\mathrm{rank}(\mathbf{W}_{\boldsymbol{\rho}})\le1$.

For perfect CSI with $\varepsilon_{\textrm{e}} = 0$, $\overline{\textrm{C7}}$ is a convex constraint that can be easily applied in SDP. On the other hand, when $\varepsilon_{\textrm{e}}>0$, $\overline{\textrm{C7}}$ becomes semi-infinite, i.e., it represents infinitely many LMIs of $\mathbf{W}_{\boldsymbol{\rho}}$ and $\mathbf{V}$, due to the continuous uncertainty set  $\mathcal{U}_{\textrm{e}}$. Hence, although the constraint is convex, the optimization problem is still computationally infeasible. To resolve this problem, transforming the semi-infinite constraint into a finite number of convex constraints is necessary and will be accomplished by the following lemma.
\vspace{-0.2cm}
\begin{lem}[Robust Quadratic Matrix Inequality \cite{Luo04quadratic}]
\label{lem:S-Lemma}\emph{ Let $f(\mathbf{X})=\mathbf{X}^{H}\mathbf{AX}+\mathbf{X}^{H}\mathbf{B}+\mathbf{B}^{H}\mathbf{X}+\mathbf{C}$, and $\mathbf{D}\succeq\mathbf{0}$. Then, the following two statements
are equivalent:}

\emph{(i) $f(\mathbf{X})\succeq\mathbf{0}$ holds for any $\mathbf{X}\in\left\{ \mathbf{X}\mid\mathrm{tr}\left(\mathbf{X}^{H}\mathbf{D}\mathbf{X}\right) \le 1\right\} $;}

\emph{(ii) There exist some $\delta \ge 0$ satisfying the following LMI, 
\begin{equation}
\left[\begin{array}{cc}
\mathbf{C}-\delta\mathbf{I} & \mathbf{B}^{H}\\
\mathbf{B} & \mathbf{A}+\delta\mathbf{D}
\end{array}\right]\succeq\mathbf{0}.
\end{equation}
}
\end{lem}
Substituting $\mathbf{G}=\widehat{\mathbf{G}}+\Delta\mathbf{G}$, constraint $\overline{\textrm{C7}}$ is equivalently reformulated as
\begin{alignat}{1}
&\Delta\mathbf{G}^{H}\mathbf{T}_{\textrm{e},\boldsymbol{\rho}}\Delta\mathbf{G}+\widehat{\mathbf{G}}^{H}\mathbf{T}_{\textrm{e},\boldsymbol{\rho}}\Delta\mathbf{G}+\Delta\mathbf{G}^{H}\mathbf{T}_{\textrm{e},\boldsymbol{\rho}}\widehat{\mathbf{G}}   \label{eq30}
 \\ 
& \qquad  +\widehat{\mathbf{G}}^{H}\mathbf{T}_{\textrm{e},\boldsymbol{\rho}}\widehat{\mathbf{G}} - \kappa_{\boldsymbol{\rho}}^{\mathrm{tol}} \mathbf{I}_{N_{\textrm{e}}} \preceq \mathbf{0},\quad \forall \Delta\mathbf{G} \in \mathcal{U}_{\textrm{e}}, \nonumber
\end{alignat}
where $\mathbf{T}_{\textrm{e},\boldsymbol{\rho}} \triangleq \mathbf{W}_{\boldsymbol{\rho}} - \frac{\kappa_{\boldsymbol{\rho}}^{\mathrm{tol}}}{\sigma_{\textrm{e}}^{2}} \mathbf{V}$. Based on Lemma \ref{lem:S-Lemma}, constraint \eqref{eq30} is equivalent to 
\vspace{-0.2cm}
\begin{alignat}{1} 
&\bigg[\begin{array}{cc}
\widehat{\mathbf{G}}^{H}\mathbf{\mathbf{T}}_{\textrm{e},\boldsymbol{\rho}}\widehat{\mathbf{G}} + ( \delta_{\textrm{e}} -  \kappa_{\boldsymbol{\rho}}^{\mathrm{tol}})\mathbf{I}_{N_{\textrm{e}}} & \widehat{\mathbf{G}}^{H}\mathbf{\mathbf{T}}_{\textrm{e},\boldsymbol{\rho}}\\
\mathbf{\mathbf{T}}_{\textrm{e},\boldsymbol{\rho}}\widehat{\mathbf{G}} & \mathbf{\mathbf{T}}_{\textrm{e},\boldsymbol{\rho}} - \frac{\delta_{\textrm{e}}}{\varepsilon_{\textrm{e}}^2}\mathbf{I}_{MN_{\mathrm{t}}}
\end{array}\bigg]\preceq\mathbf{0}, \nonumber \\
& \qquad \qquad \qquad \exists \delta_{\textrm{e}}\ge0, \nonumber \\
&  \iff\widetilde{\textrm{C7}}\textrm{: }\mathbf{U}_{\textrm{e}}^{H}\mathbf{\mathbf{T}}_{\textrm{e},\boldsymbol{\rho}}\mathbf{U}_{\textrm{e}} \preceq \bigg[\begin{array}{cc}
(\kappa_{\boldsymbol{\rho}}^{\mathrm{tol}} - \delta_{\textrm{e}})\mathbf{I}_{N_{\textrm{e}}} & \mathbf{0}\\
\mathbf{0} & \frac{\delta_{\textrm{e}}}{\varepsilon_{\textrm{e}}^2}\mathbf{I}_{MN_{\mathrm{t}}}
\end{array}\bigg],  \nonumber \\
& \qquad \qquad \qquad \delta_{\textrm{e}} \ge 0. 
\end{alignat}
Besides, constraints C7 and $\widetilde{\textrm{C7}}$ are equivalent if $\mathrm{rank}(\mathbf{W}_{\boldsymbol{\rho}})\le1$. This completes the proof. 

\section{\label{append2}Proof of Theorem \ref{prop2}}
Note that problems R0($\mathbf{D}_{\text{\mbox{II}},2}$) and R1 are equivalent if and only if the rank constraint $\mathrm{rank}(\mathbf{W}_{\boldsymbol{\rho}}^{*})\le1$ is fulfilled. Below we only prove the result for problem R1 with $\varepsilon_{\textrm{e}}\!>\!0$ and $\widetilde{\textrm{C7}}$. The result for $\varepsilon_{\textrm{e}}\!=\!0$ and $\overline{\textrm{C7}}$ can be proved similarly; see \cite{Xiang16:CoMP}.

Let $\boldsymbol{\alpha}=[\alpha_{m\boldsymbol{\rho}}]$, $\boldsymbol{\beta}=[\beta_{m}]$, $\boldsymbol{\lambda}=[\lambda_{\boldsymbol{\rho}}]$, $\boldsymbol{\Phi}_{\boldsymbol{\rho}}$, and $\boldsymbol{\Theta}_{\boldsymbol{\rho}}=[\boldsymbol{\Theta}_{1\boldsymbol{\rho}},\,\boldsymbol{\Theta}_{2\boldsymbol{\rho}}]$ be the Lagrangian multipliers associated with constraints $\overline{\textrm{C4}}$, $\overline{\textrm{C5}}$, $\overline{\textrm{C6}}$, $\widetilde{\textrm{C7}}$, and $\overline{\textrm{C10}}$, respectively, where 
\begin{alignat*}{1}
\alpha_{m\boldsymbol{\rho}}\ge0,\beta_{m}\ge0,\lambda_{\boldsymbol{\rho}}\ge0,\boldsymbol{\Phi}_{\boldsymbol{\rho}}\succeq\mathbf{0},\boldsymbol{\Theta}_{1\boldsymbol{\rho}}\succeq\mathbf{0},\textrm{ and }\boldsymbol{\Theta}_{2\boldsymbol{\rho}}\succeq\mathbf{0}.
\end{alignat*}
The Lagrangian of problem R1 is  
\begin{alignat}{1}
\mathcal{L}(\mathbf{W}_{\boldsymbol{\rho}}, \! \mathbf{V};\!\boldsymbol{\Upsilon}) &=  \sum_{\boldsymbol{\rho}} \! \mathrm{tr} \! \big[\big(\mathbf{B}_{1\boldsymbol{\rho}} - 2 \lambda_{\boldsymbol{\rho}} \mathbf{H}_{\boldsymbol{\rho}} - \boldsymbol{\Theta}_{1\boldsymbol{\rho}} \big)\mathbf{W}_{\boldsymbol{\rho}}   \\ 
& + \big(\mathbf{B}_{2 \boldsymbol{\rho}} - \kappa_{\boldsymbol{\rho}}^{\mathrm{tol}} \widehat{\mathbf{G}} \boldsymbol{\Phi}_{\boldsymbol{\rho}} \widehat{\mathbf{G}}^{H}-\boldsymbol{\Theta}_{2\boldsymbol{\rho}} \big) \mathbf{V} \big] + \Delta_0, \nonumber
\end{alignat}
where $\boldsymbol{\Upsilon} \triangleq [\boldsymbol{\alpha},\boldsymbol{\beta},\boldsymbol{\lambda},\boldsymbol{\Phi}_{\boldsymbol{\rho}},\boldsymbol{\Theta}_{\boldsymbol{\rho}}]$, $\Delta_0$ is a collection of terms irrelevant for the proof, and 
\begin{alignat}{1}
\mathbf{B}_{1\boldsymbol{\rho}} & \triangleq\mathbf{I}+\boldsymbol{\Lambda}_{\boldsymbol{\rho}}^{\boldsymbol{\alpha},\boldsymbol{\beta}}+\widehat{\mathbf{G}}\boldsymbol{\Phi}_{\boldsymbol{\rho}}\widehat{\mathbf{G}}^{H}+\sum_{\boldsymbol{\rho}\in\mathcal{S}}(1+\kappa_{\boldsymbol{\rho}}^{\mathrm{req}})\lambda_{\boldsymbol{\rho}}\mathbf{H}_{\boldsymbol{\rho}}\succ\mathbf{0}, \nonumber \\
\mathbf{B}_{2\boldsymbol{\rho}} & \triangleq\mathbf{I}+\boldsymbol{\Lambda}^{\boldsymbol{\beta}}+\lambda_{\boldsymbol{\rho}}\mathbf{H}_{\boldsymbol{\rho}}\succ\mathbf{0}, 
\end{alignat}
with $\boldsymbol{\Lambda}_{\boldsymbol{\rho}}^{\boldsymbol{\alpha},\boldsymbol{\beta}}\triangleq\sum_{m\in\mathcal{M}}(\alpha_{m\boldsymbol{\rho}}+\beta_{m})\boldsymbol{\Lambda}_{m}$
and $\boldsymbol{\Lambda}^{\boldsymbol{\beta}}\triangleq\sum_{m\in\mathcal{M}}\beta_{m}\boldsymbol{\Lambda}_{m}$.
It can be verified that R1 is a convex optimization problem and fulfills Slater's constraint qualification. Thus, strong duality holds for problem R1 and the Karush\textendash Kuhn\textendash Tucker (KKT) conditions are both necessary and sufficient for a primal-dual point $(\mathbf{W}_{\boldsymbol{\rho}},\mathbf{V};\,\boldsymbol{\Upsilon})$ to be optimal. The KKT conditions for problem R1 are given by
\vspace{-.2cm} 
\begin{alignat}{1}
\nabla_{\mathbf{W}_{\boldsymbol{\rho}}}\mathcal{L}=\mathbf{B}_{1\boldsymbol{\rho}}-2\lambda_{\boldsymbol{\rho}}\mathbf{H}_{\boldsymbol{\rho}}-\boldsymbol{\Theta}_{1\boldsymbol{\rho}}=\mathbf{0},\label{eq:kkt1}\\
\mathbf{W}_{\boldsymbol{\rho}}\succeq\mathbf{0},\quad\lambda_{\boldsymbol{\rho}}\ge0, \quad \mathbf{W}_{\boldsymbol{\rho}}\boldsymbol{\Theta}_{1\boldsymbol{\rho}}=\mathbf{0}.\label{eq:kkt2}
\end{alignat}
Based on \eqref{eq:kkt1} and \eqref{eq:kkt2}, we have $\mathbf{W}_{\boldsymbol{\rho}}\mathbf{B}_{1\boldsymbol{\rho}}=2\lambda_{\boldsymbol{\rho}}\mathbf{W}_{\boldsymbol{\rho}}\mathbf{H}_{\boldsymbol{\rho}}$. Besides, constraint $\overline{\textrm{C6}}$ is satisfied with equality for the optimal solution and thus $\lambda_{\boldsymbol{\rho}} > 0$. Moreover, since $\mathrm{rank}(\mathbf{H}_{\boldsymbol{\rho}}) \le1$, the rank of the optimal $\mathbf{W}_{\boldsymbol{\rho}}$ can be determined as 
\vspace{-0.1cm}
\begin{alignat}{1}
\mathrm{rank}(\mathbf{W}_{\boldsymbol{\rho}}) & \stackrel{\textrm{(a)}}{=}\mathrm{rank}(\mathbf{W}_{\boldsymbol{\rho}}\mathbf{B}_{1\boldsymbol{\rho}})\stackrel{\textrm{(b)}}{=}\mathrm{rank}(\lambda_{\boldsymbol{\rho}}\mathbf{W}_{\boldsymbol{\rho}}\mathbf{H}_{\boldsymbol{\rho}}) \\& 
 \stackrel{\textrm{(c)}}{\le}\min\left\{ \mathrm{rank}(\lambda_{\boldsymbol{\rho}}\mathbf{W}_{\boldsymbol{\rho}}),\,\mathrm{rank}(\mathbf{H}_{\boldsymbol{\rho}})\right\} \le1, \nonumber
\end{alignat}
where (a) is due to $\mathbf{B}_{1\boldsymbol{\rho}}\succ\mathbf{0}$, (b) is a result of \eqref{eq:kkt1} and \eqref{eq:kkt2}, and (c) follows from the rank inequality $\mathrm{rank}(\mathbf{AB})\le\min\left\{ \mathrm{rank}(\mathbf{A}),\,\mathrm{rank}(\mathbf{B})\right\}$ \cite{Horn2012Matrix}. Thus, $\mathrm{rank}(\mathbf{W}^*_{\boldsymbol{\rho}})\le1$ has to hold if problem R1 is feasible. This completes the proof.

\section{\label{append}Proof of Theorem~\ref{thm2}}
We first assume that the caching decisions are given in problems Q0 and Q1, where the resulting problems are denoted by Q0($\mathbf{D}_{\text{\mbox{I}},\omega}$) and Q1($\mathbf{D}_{\text{\mbox{I}},\omega}$), respectively. Without loss of generality, Q0($\mathbf{D}_{\text{\mbox{I}},\omega}$) can be written in general form as 
\vspace{-0.2cm}
\begin{align}
\textrm{Q0(\ensuremath{\mathbf{D}_{\text{\mbox{I}},\omega}}):}\;\minimize%_{\mathbf{D}_{\text{\mbox{I}},\omega}}
\;\; & \frac{1}{\Omega}\sum\nolimits _{\omega=1}^{\Omega}\; f_{\mathrm{\text{\mbox{I}}},\omega}(\mathbf{D}_{\text{\mbox{I}},\omega})\\
{\st}\;\; & \sum\nolimits _{\omega=1}^{\Omega}\mathbf{g}_{\mathrm{\text{\mbox{I}}},\omega}(\mathbf{D}_{\text{\mbox{I}},\omega})\preceq\mathbf{0}, \nonumber \\
&\mathbf{D}_{\text{\mbox{I}},\omega}\in\mathbf{\mathcal{D}}_{\text{\mbox{I}},\omega},\;\omega\in\left\{ 1,\ldots,\Omega\right\} ,\nonumber  \\
{\revise{\var}} \;\; & {\revise{\mathbf{D}_{\text{\mbox{I}},\omega} = [q_{f,l,m,\omega},b_{f,l,m,\omega},\mathbf{w}_{\boldsymbol{\rho},\omega},\mathbf{V}_{\omega}], }} \nonumber
\end{align}
where $\mathbf{g}_{\mathrm{\text{\mbox{I}}},\omega} \triangleq [g_{\mathrm{\text{\mbox{I}}},\omega,1},\ldots,g_{\mathrm{\text{\mbox{I}}},\omega,M}] \textrm{: }\mathbb{R}^{\Omega \times 1} \to\mathbb{R}^{M \times 1}$ is an affine vector-valued function and represents the backhaul constraint $\ensuremath{\overline{\textrm{C3}}}$ with $ {g}_{\mathrm{\text{\mbox{I}}}, \omega,m} (\cdot) \triangleq \sum\nolimits _{f\in\mathcal{F}}b_{f,l,m,\omega}Q_{f}  - B_{m,\omega}^{\max}$. Herein,  $\mathbf{\mathcal{D}}_{\text{\mbox{I}}, \omega}$ is a non-convex set because of C2. We assume that Q0($\mathbf{D}_{\text{\mbox{I}},\omega}$) is feasible. Let $f^{*}$ and $q^{*}$ denote the primal and the dual optimal values of Q0($\mathbf{D}_{\text{\mbox{I}},\omega}$), respectively.

Meanwhile, Q1($\mathbf{D}_{\text{\mbox{I}},\omega}$) is obtained from Q0($\mathbf{D}_{\text{\mbox{I}},\omega}$) by relaxing the binary constraint C2. Due to the convexity of Q1, strong duality holds for problem Q1($\mathbf{D}_{\text{\mbox{I}},\omega}$). Based on Lagrangian duality theory, it can be further shown that the dual problems of Q1($\mathbf{D}_{\text{\mbox{I}},\omega}$) and Q0($\mathbf{D}_{\text{\mbox{I}},\omega}$) are identical \cite[Chapter 5.5.3]{Bertsekas99NLP}. Consequently, the optimal value of Q1($\mathbf{D}_{\text{\mbox{I}},\omega}$) is also given by $q^{*}$. Then, Theorem~\ref{thm2} can be proved by resorting to the following proposition, which estimates the duality gap for   Q0($\mathbf{D}_{\text{\mbox{I}},\omega}$).
%\vspace{-0.3cm}
\begin{prop}
\label{propa}
\emph{\label{prop-duality-gap} For problem Q0($\mathbf{D}_{\text{\mbox{I}},\omega}$), the duality gap, $f^* - q^*$, is bounded and satisfies   
\begin{equation}
0\le f^{*}-q^{*}\le\mathcal{O}\left(\frac{M+1}{\Omega}\right).\label{eq-duality-gap}
\end{equation}
}\end{prop}
\vspace{-0.1cm} 

Specifically, according to \eqref{eq-duality-gap}, the difference between the optimal values of Q1($\mathbf{D}_{\text{\mbox{I}},\omega}$) and Q0($\mathbf{D}_{\text{\mbox{I}},\omega}$) becomes negligible for a sufficiently large value of $\Omega$,
i.e., 
\vspace{-0.2cm}
\begin{equation}
\lim_{\Omega\to\infty}(f^{*}-q^{*})=0.\label{eq:nullgap}
\end{equation}
Since \eqref{eq:nullgap} holds for arbitrary caching decisions, the performance gap between Q1 and Q0 also vanishes as $\Omega\to\infty$. Therefore, the remainder of the proof will be focused on establishing \eqref{eq-duality-gap} in Proposition~\ref{propa}.  

\begin{IEEEproof}[Proof of Proposition~\ref{propa}]
The left hand side inequality in \eqref{eq-duality-gap} is simply due to the weak duality property for general nonlinear optimization problems \cite[Chapter 5.1.2]{Bertsekas99NLP}. To prove the right hand side of \eqref{eq-duality-gap}, let us define the sets $\mathcal{X}_{\omega}\triangleq\big\{ \mathbf{x}_{\omega}\triangleq[\mathbf{g}_{\mathrm{\text{\mbox{I}}},\omega}(\mathbf{D}_{\text{\mbox{I}},\omega}),\, f_{\mathrm{\text{\mbox{I}}},\omega}(\mathbf{D}_{\text{\mbox{I}},\omega})]\in\mathbb{R}^{M+1}\mid\mathbf{D}_{\text{\mbox{I}},\omega}\in\mathbf{\mathcal{D}}_{\text{\mbox{I}},\omega}\big\} ,\;\omega\in\left\{ 1,\ldots,\Omega\right\} $,
and their vector (Minkowski) sum $\mathcal{X} \! \triangleq \! \left\{ \mathbf{x} \! = \! \sum_{\omega=1}^{\Omega}\mathbf{x}_{\omega}\mid\mathbf{x}_{\omega} \in\mathcal{X}_{\omega}\right\} \! \subseteq \! {\mathbb{R}^{M+1}}$.
Here, $\mathbf{x}_{\omega}$ defines an achievable constraint-objective value pair. 
For simplicity of notation, we also write $\mathcal{X} = \sum_{\omega=1}^{\Omega}\mathcal{X}_{\omega}$. Using $\mathcal{X}$ and its convex hull, $\mathrm{conv}(\mathcal{X})$, the primal optimum and the dual optimum of Q0($\mathbf{D}_{\text{\mbox{I}},\omega}$) are given by 
\vspace{-0.2cm}
\begin{alignat}{1}
f^{*} &=  \min\left\{ z\mid(\mathbf{y},z)\in\mathcal{X}\textrm{ with }\mathbf{y}\preceq\mathbf{0}\right\}  \; \textrm{and} \label{aprimalopt}
 \\
q^{*} & = \min\left\{ z\mid(\mathbf{y},z)\in\mathrm{conv}(\mathcal{X})\textrm{ with }\mathbf{y}\preceq\mathbf{0}\right\}, \nonumber
\vspace{-0.1cm}
\end{alignat}
respectively. Assume that $(\mathbf{y}^{*},q^{*}) \in \mathrm{conv}(\mathcal{X})$ obtains the dual optimum with $\mathbf{y}^{*}\preceq\mathbf{0}$.

The estimation of the duality gap is feasible due to the Shapley-Folkman theorem \cite{Aubin76:DualGap},\cite[Proposition 5.26]{Bertsekas82Lagrange}. Specifically, constrained by the dimension of its subspace $\mathbb{R}^{(M+1)\times 1}$, each point of set $\mathrm{conv}(\mathcal{X}) \subseteq \mathbb{R}^{(M+1) \times 1} $ can be represented as the vector sum of at least $(\Omega - M - 1)$ out of $\Omega$ points in $\mathcal{X}_{\omega}$. This means, for each $(\mathbf{y}^{*},q^{*}) \! \in\! \mathrm{conv}(\mathcal{X}) $, there exist two  index subsets $\mathcal{I}, \overline{\mathcal{I}} \subset \left\{ 1,\ldots,\Omega\right\} $ satisfying $\mathcal{I} \cap \overline{\mathcal{I}} = \emptyset$, $\mathcal{I} \cup \overline{\mathcal{I}} = \left\{ 1,\ldots,\Omega\right\}$, and  $\left|\mathcal{I}\right|\le M+1$, such that  
\vspace{-0.2cm}
\begin{alignat}{1}
&\sum\nolimits_{\omega\in\mathcal{I}}\mathbf{y}_{\omega}^{*}+ \sum\nolimits_{\omega \in \overline{\mathcal{I}}}\mathbf{g}_{\mathrm{\text{\mbox{I}}},\omega}(\overline{\mathbf{D}}_{\text{\mbox{I}},\omega}) \preceq\mathbf{0}\quad\textrm{and} \label{aeq46}
 \\
 & \sum\nolimits_{\omega\in\mathcal{I}}q_{\omega}^{*} + \sum\nolimits_{\omega \in \overline{\mathcal{I}}}f_{\mathrm{\text{\mbox{I}}},\omega}(\overline{\mathbf{D}}_{\text{\mbox{I}},\omega})  =q^{*}\Omega  \nonumber
\end{alignat}
hold for $(\mathbf{y}_{\omega}^{*},q_{\omega}^{*})\in\mathrm{conv}(\mathcal{X}_{\omega})$,
$\omega\in\mathcal{I}$ and $\overline{\mathbf{D}}_{\text{\mbox{I}},\omega}\in\mathbf{\mathcal{D}}_{\text{\mbox{I}},\omega}$, $\omega \in \overline{\mathcal{I}}$. In other words, at most $M+1$ vectors obtained from the dual (relaxed) problem can be infeasible for the primal problem Q0($\mathbf{D}_{\text{\mbox{I}},\omega}$), i.e., $(\mathbf{y}_{\omega}^{*},q_{\omega}^{*})\in\mathrm{conv}(\mathcal{X}_{\omega})$ but $(\mathbf{y}_{\omega}^{*},q_{\omega}^{*})\notin\mathcal{X}_{\omega}$ for $\omega\in\mathcal{I}$.

Therefore, we construct suboptimal solutions for Q0($\mathbf{D}_{\text{\mbox{I}},\omega}$) from the dual (relaxed) solutions in \eqref{aeq46}. Let $\alpha_{\omega}^{i}\in[0,1]$ and $\sum_{i=1}^{M+2}\alpha_{\omega}^{i}=1$. For each $\omega\in\mathcal{I}$, we can express $\mathbf{y}_{\omega}^{*}$ and $q_{\omega}^{*}$ as convex combinations of $\hat{\mathbf{D}}_{\text{\mbox{I}},\omega}^{i}\in\mathbf{\mathcal{D}}_{\text{\mbox{I}},\omega}$, $i\in\left\{ 1,\ldots,M+2\right\} $, i.e., 
\begin{alignat}{1}
\mathbf{y}_{\omega}^{*} & =\sum\nolimits _{i=1}^{M+2}\alpha_{\omega}^{i}\mathbf{g}_{\mathrm{\text{\mbox{I}}},\omega}(\hat{\mathbf{D}}_{\text{\mbox{I}},\omega}^{i}) \label{aeq45}
 \\ %,  \; \mathbf{y}_{\omega}^{*}  
&\succeq \overline{\mathbf{g}}_{\mathrm{\text{\mbox{I}}},\omega}\triangleq\min\Big\{ \sum\nolimits_{i=1}^{M+2}\alpha_{\omega}^{i}\mathbf{g}_{\mathrm{\text{\mbox{I}}},\omega}(\hat{\mathbf{D}}_{\text{\mbox{I}},\omega}^{i})\mid\hat{\mathbf{D}}_{\text{\mbox{I}},\omega}^{i}\in\mathbf{\mathcal{D}}_{\text{\mbox{I}},\omega}\Big\}, \nonumber \\ %\; \textrm{and} \nonumber \\
q_{\omega}^{*} & =\sum\nolimits _{i=1}^{M+2} \alpha_{\omega}^{i} f_{\mathrm{\text{\mbox{I}}},\omega}(\hat{\mathbf{D}}_{\text{\mbox{I}},\omega}^{i}) \nonumber \\ %, \; q_{\omega}^{*} 
&\ge \overline{f}_{\mathrm{\text{\mbox{I}}},\omega} \triangleq \min\left\{ \sum\nolimits_{i=1}^{M+2}\alpha_{\omega}^{i}f_{\mathrm{\text{\mbox{I}}},\omega}(\hat{\mathbf{D}}_{\text{\mbox{I}},\omega}^{i})\mid\hat{\mathbf{D}}_{\text{\mbox{I}},\omega}^{i}\in\mathbf{\mathcal{D}}_{\text{\mbox{I}},\omega}\right\} . \nonumber
\end{alignat}
We define the primal solution $\overline{\mathbf{D}}_{\text{\mbox{I}},\omega} \in \mathbf{\mathcal{D}}_{\text{\mbox{I}}, \omega}$, $\omega \in \mathcal{I} \cup \overline{\mathcal{I}} $, with  
\vspace{-0.2cm}
\begin{alignat}{1}
\overline{\mathbf{D}}_{\text{\mbox{I}},\omega}  \in & \, {\argmin} \big\{ f_{\mathrm{\text{\mbox{I}}},\omega}(\mathbf{D}_{\text{\mbox{I}},\omega})\mid\mathbf{g}_{\mathrm{\text{\mbox{I}}},\omega}(\mathbf{D}_{\text{\mbox{I}},\omega})\preceq \overline{\mathbf{g}}_{\mathrm{\text{\mbox{I}}},\omega}, \label{eq46} \\
& \qquad \qquad \mathbf{D}_{\text{\mbox{I}},\omega}\in\mathbf{\mathcal{D}}_{\text{\mbox{I}},\omega}\big\} , \nonumber
\end{alignat}
for $\omega\in\mathcal{I}$. The feasible set of \eqref{eq46} is generally nonempty and thus $\overline{\mathbf{D}}_{\text{\mbox{I}},\omega}$ usually exists. Meanwhile, $\overline{\mathbf{D}}_{\text{\mbox{I}},\omega}$ is primal feasible since $\mathbf{g}_{\mathrm{\text{\mbox{I}}},\omega}(\overline{\mathbf{D}}_{\text{\mbox{I}},\omega}) \! \preceq \! \mathbf{y}_{\omega}^{*} \! \preceq \! \mathbf{0}$, but generally suboptimal for \eqref{aprimalopt} since $ \sum\nolimits_{\omega \in \mathcal{I} \cup \overline{\mathcal{I}}} f_{\mathrm{\text{\mbox{I}}},\omega}(\overline{\mathbf{D}}_{\text{\mbox{I}},\omega})\ge f^{*}\Omega$.

However, $\overline{\mathbf{D}}_{\text{\mbox{I}},\omega}$ incurs only a bounded penalty on the objective value,
\vspace{-0.2cm}
\begin{equation}
f_{\mathrm{\text{\mbox{I}}},\omega}(\overline{\mathbf{D}}_{\text{\mbox{I}},\omega}) - q_{\omega}^* \overset{\textrm{(a)}}{\le}  f_{\mathrm{\text{\mbox{I}}},\omega}(\overline{\mathbf{D}}_{\text{\mbox{I}},\omega}) - \overline{f}_{\mathrm{\text{\mbox{I}}},\omega} \le \varrho_{\omega} \overset{\textrm{(b)}}{\le}  f_{\mathrm{\text{\mbox{I}}},\omega}^{UB} - f_{\mathrm{\text{\mbox{I}}},\omega}^{LB},
\label{fsubopt}
\end{equation}
where $f_{\mathrm{\text{\mbox{I}}},\omega}^{UB} \triangleq \max\left\{ f_{\mathrm{\text{\mbox{I}}},\omega}(\mathbf{D}_{\text{\mbox{I}},\omega}) \mid\mathbf{D}_{\text{\mbox{I}},\omega}\in\mathbf{\mathcal{D}}_{\text{\mbox{I}},\omega}\right\}$ and $f_{\mathrm{\text{\mbox{I}}},\omega}^{LB} \triangleq \min\left\{ f_{\mathrm{\text{\mbox{I}}},\omega}(\mathbf{D}_{\text{\mbox{I}},\omega})\mid\mathbf{D}_{\text{\mbox{I}},\omega}\in\mathbf{\mathcal{D}}_{\text{\mbox{I}},\omega}\right\}$. In \eqref{fsubopt}, (a) is due to \eqref{aeq45}, and (b) holds  since $f_{\mathrm{\text{\mbox{I}}},\omega}^{UB} \ge f_{\mathrm{\text{\mbox{I}}},\omega}({\mathbf{D}}_{\text{\mbox{I}},\omega}) \ge  f_{\mathrm{\text{\mbox{I}}},\omega}^{LB}$, $\forall \mathbf{D}_{\text{\mbox{I}},\omega}\in\mathbf{\mathcal{D}}_{\mathrm{\text{\mbox{I}}},\omega}$.
We have $\varrho_{\omega}<+\infty$ since $f_{\mathrm{\text{\mbox{I}}},\omega}(\cdot)$ is a continuous function and $\mathbf{\mathcal{D}}_{\text{\mbox{I}},\omega}\neq\emptyset$ if problem Q0($\mathbf{D}_{\text{\mbox{I}},\omega}$) is feasible. 
Let $\varrho_{\max}\triangleq\max\left\{ \varrho_{\omega}\mid\omega=1,\ldots,\Omega\right\} $. The duality gap finally satisfies 
\begin{alignat}{1}
f^* & \le \frac{1}{\Omega} \sum\nolimits_{\omega \in \mathcal{I} \cup \overline{\mathcal{I}}} f_{\mathrm{\text{\mbox{I}}},\omega}(\mathbf{D}_{\text{\mbox{I}},\omega})  \\
& \overset{\textrm{(a)}}{\le} q^* + \frac{1}{\Omega} \sum\nolimits_{\omega \in \mathcal{I}} \left(  \overline{f}_{\mathrm{\text{\mbox{I}}},\omega} - q_{\omega}^* +  \varrho_{\omega}  \right)  \nonumber \\
& \le q^* + \frac{M+1}{\Omega}\varrho_{\max}, \nonumber
\end{alignat}
where (a) is due to \eqref{aeq46} and \eqref{fsubopt}. This completes the proof. 
\end{IEEEproof}
\vspace{-.2cm}
\begin{rem}
Eq. \eqref{eq:nullgap} implies that the vector sum of $\Omega$ sets, $\mathcal{X}$, in subspace $\mathbb{R}^{(M+1)\times 1}$ tends to be convexified as $\Omega \to \infty$ or $M\ll \Omega$, in the sense that any vector in its convex hull, $\mathrm{conv} (\mathcal{X})$, can be closely approximated by a vector in $\mathcal{X}$ itself due to the underlying geometry. This property has been exploited to solve MINLPs in several disciplines \cite{Bertsekas83:UC,Giannakis13:SG}. %Nikola13JSAC:BaR 
\end{rem}
\vspace{-0.2cm}

%\bibliographystyle{IEEEtran}
%\bibliography{IEEEabrv,bib/SecureCaching,bib/MIMO-IBC,bib/PLS,bib/OptimizationRefs,bib/WirelessCaching,bib/CoMP,bib/WirelessVideo,bib/SecureCachingRevision}

\begin{thebibliography}{10}


\bibitem{Xiang16:CoMP}
L.~Xiang, D.~W.~K. Ng, R.~Schober, and V.~W.~S. Wong, ``Cache-enabled
  physical-layer security for video streaming in backhaul-limited cellular
  networks,'' in \emph{Proc. IEEE Global Comm. Conf. (GLOBECOM) - Workshop on
  Trusted Communications with Physical Layer Security}, Washington, DC, Dec.
  2016.

\bibitem{WSJ13:Verizon}
R.~Knutson, ``Video boom forces {Verizon} to upgrade network,'' \emph{The Wall
  Street Journal}, Dec. 2013.

\bibitem{Rost14MCOM:CloudRAN}
P.~Rost, C.~Bernardos, A.~Domenico, M.~Girolamo, M.~Lalam, A.~Maeder,
  D.~Sabella \emph{et~al.}, ``Cloud technologies for flexible {5G} radio access
  networks,'' \emph{{IEEE} Commun. Mag.}, vol.~52, no.~5, pp. 68--76, May 2014.

\bibitem{Paschos16WC}
G.~Paschos, E.~Ba{\c{s}}tu{\u{g}}, I.~Land, G.~Caire, and M.~Debbah, ``Wireless
  caching: {T}echnical misconceptions and business barriers,'' \emph{{IEEE}
  Commun. Mag.}, pp. 16--22, Aug. 2016.

\bibitem{Liu13TSP:CoMP}
A.~Liu and V.~Lau, ``Mixed-timescale precoding and cache control in cached
  {MIMO} interference network,'' \emph{{IEEE} Trans. Signal Process.}, vol.~61,
  no.~24, pp. 6320--6332, Dec. 2013.

\bibitem{Liu14TSP:CoMP}
------, ``Cache-enabled opportunistic cooperative {MIMO} for video streaming in
  wireless systems,'' \emph{{IEEE} Trans. Signal Process.}, vol.~62, no.~2, pp.
  390--402, Jan. 2014.

\bibitem{chen16cooperative}
Z.~Chen, J.~Lee, T.~Q. Quek, and M.~Kountouris, ``Cooperative caching and
  transmission design in cluster-centric small cell networks,'' in \emph{IEEE Trans. Wireless Commun.}, vol. 16, no. 5, pp. 3401-3415, May 2017.

\bibitem{Tao16TWC:Multicast}
M.~Tao, E.~Chen, H.~Zhou, and W.~Yu, ``Content-centric sparse multicast
  beamforming for cache-enabled cloud {RAN},'' \emph{{IEEE} Trans. Wireless
  Commun.}, vol.~15, no.~9, pp. 6118--6131, Sep. 2016.

\bibitem{Xiang17TVT:CLCaching}
L.~Xiang, D.~W.~K. Ng, T.~Islam, R.~Schober, V.~W.~S. Wong, and J.~Wang,
  ``Cross-layer optimization of fast video delivery in cache- and
  buffer-enabled relaying networks,'' \emph{{IEEE} Trans. Veh. Technol.},
  vol.~PP, no.~99, 2017.

\bibitem{Breslau99Zipf}
L.~Breslau, P.~Cao, L.~Fan, G.~Phillips, and S.~Shenker, ``Web caching and
  {Z}ipf-like distributions: Evidence and implications,'' in \emph{Proc. IEEE
  INFOCOM}, New York, NY, Mar. 1999.

\bibitem{Peng16FogRAN}
M.~Peng, S.~Yan, K.~Zhang, and C.~Wang, ``Fog-computing-based radio access
  networks: {Issues} and challenges,'' \emph{{IEEE} Netw.}, vol.~30, no.~4, pp.
  46--53, Jul.-Aug. 2016.

\bibitem{Park16TWC:FogRAN}
S.-H. Park, O.~Simeone, and S.~S. Shitz, ``Joint optimization of cloud and edge
  processing for fog radio access networks,'' \emph{{IEEE} Trans. Wireless
  Commun.}, vol.~15, no.~11, pp. 7621--7632, Nov. 2016.

\bibitem{Tandon16GC}
S.~M. Azimi, O.~Simeone, and R.~Tandon, ``Fundamental limits on latency in
  small-cell caching systems: {An} information-theoretic analysis,'' in
  \emph{Proc. IEEE Global Comm. Conf. (GLOBECOM)}, Washington, DC, Dec. 2016.

\bibitem{Tandon17online}
S.~M. Azimi, O.~Simeone, A.~Sengupta, and R.~Tandon, ``Online edge caching in
  fog-aided wireless network,'' in \emph{Proc. IEEE Int. Sym. Inf. Theory (ISIT)}, Aachen, Germany, Jun.
  2017.

\bibitem{Clancy15IT:Limits}
A.~Sengupta, R.~Tandon, and T.~C. Clancy, ``Fundamental limits of caching with
  secure delivery,'' \emph{{IEEE} Trans. Inf. Forensics Security}, vol.~10,
  no.~2, pp. 355--370, Feb. 2015.

\bibitem{AwanICC15:D2D}
Z.~H. Awan and A.~Sezgin, ``Fundamental limits of caching in {D2D} networks
  with secure delivery,'' in \emph{Proc. IEEE Int. Conf. Comm. (ICC) -
  Workshop on Wireless Physical Layer Security}, London, UK, Jun. 2015.

\bibitem{LandICC16:HetNets}
F.~Gabry, V.~Bioglio, and I.~Land, ``On edging caching with secrecy
  constraints,'' in \emph{Proc. IEEE Int. Conf. Comm. (ICC)}, Kuala Lumpur,
  Malaysia, May 2016.

\bibitem{Niesen14IT:CodedCaching}
M.~A. Maddah-Ali and U.~Niesen, ``Fundamental limits of caching,'' \emph{{IEEE}
  Trans. Inf. Theory}, vol.~60, no.~5, pp. 2856--2867, May 2014.

\bibitem{Khisti10IT:MISOME}
A.~Khisti and G.~W. Wornell, ``Secure transmission with multiple antennas {I}:
  The {MISOME} wiretap channel,'' \emph{{IEEE} Trans. Inf. Theory}, vol.~56,
  no.~7, pp. 3088--3104, Jul. 2010.

\bibitem{Khisti10IT:MIMOME}
------, ``Secure transmission with multiple antennas {II}: The {MIMOME} wiretap
  channel,'' \emph{{IEEE} Trans. Inf. Theory}, vol.~56, no.~11, pp. 5515--5532,
  Nov. 2010.

\bibitem{Liu15TWC:SDOF}
T.-Y. Liu, P.~Mukherjee, S.~Ulukus, S.-C. Lin, and Y.-W.~P. Hong, ``Secure
  degrees of freedom of {MIMO} {R}ayleigh block fading wiretap channels with no
  {CSI} anywhere,'' \emph{{IEEE} Trans. Wireless Commun.}, vol.~14, no.~5, pp.
  2655--2669, May 2015.

\bibitem{Jafar16IT}
A.~G. Davoodi and S.~A. Jafar, ``Aligned image sets under channel uncertainty:
  {Settling} conjectures on the collapse of degrees of freedom under finite
  precision {CSIT},'' \emph{{IEEE} Trans. Inf. Theory}, vol.~62, no.~10, pp.
  5603--5618, Oct. 2016.

\bibitem{Gesbert10JSAC:CoMP}
D.~Gesbert, S.~Hanly, H.~Huang, S.~S. Shitz, O.~Simeone, and W.~Yu,
  ``Multi-cell {MIMO} cooperative networks: {A} new look at interference,''
  \emph{{IEEE} J. Sel. Areas Commun.}, vol.~28, no.~9, pp. 1380--1408, Dec.
  2010.

\bibitem{Negi08TWC:AN}
S.~Goel and R.~Negi, ``Guaranteeing secrecy using artificial noise,''
  \emph{{IEEE} Trans. Wireless Commun.}, vol.~7, no.~6, pp. 2180--2189, Jun.
  2008.

\bibitem{LiTSP11:Opt-Robust}
Q.~Li and W.-K. Ma, ``Optimal and robust transmit designs for {MISO} channel
  secrecy by semidefinite programming,'' \emph{{IEEE} Trans. Signal Process.},
  vol.~59, no.~8, pp. 3799--3812, Aug. 2011.

\bibitem{Ng15TWC}
D.~W.~K. Ng and R.~Schober, ``Secure and green {SWIPT} in distributed antenna
  networks with limited backhaul capacity,'' \emph{{IEEE} Trans. Wireless
  Commun.}, vol.~14, no.~9, pp. 5082--5097, Sep. 2015.

\bibitem{Ma11:MobVid}
K.~J. Ma, R.~Bartos, S.~Bhatia, and R.~Nair, ``Mobile video delivery with
  {HTTP},'' \emph{{IEEE} Commun. Mag.}, vol.~49, no.~4, pp. 166--175, Apr.
  2011.

\bibitem{ICC14:Learning}
P.~Blasco and D.~G\"{u}nd\"{u}z, ``Learning-based optimization of cache content in a
  small cell base station,'' in \emph{Proc. IEEE Int. Conf. Comm. (ICC)},
  Sydney, Australia, Jun. 2014.

\bibitem{Birge2011IntroSP}
J.~R. Birge and F.~Louveaux, \emph{Introduction to {S}tochastic
  {P}rogramming}.\hskip 1em plus 0.5em minus 0.4em\relax Springer Science \&
  Business Media, 2011.

\bibitem{Shapiro2009LectSP}
A.~Shapiro, D.~Dentcheva, and A.~Ruszczy\'{n}ski, \emph{Lectures on
  {S}tochastic {P}rogramming: {M}odeling and {T}heory}.\hskip 1em plus 0.5em
  minus 0.4em\relax SIAM, 2009.

\bibitem{Tse2005Fundamentals}
D.~Tse and P.~Viswanath, \emph{Fundamentals of {W}ireless
  {C}ommunication}.\hskip 1em plus 0.5em minus 0.4em\relax Cambridge University
  Press, 2005.

\bibitem{Cover12IT}
T.~M. Cover and J.~A. Thomas, \emph{Elements of {Information} {Theory}},  2nd~ed.\hskip 1em plus 0.5em minus 0.4em\relax John Wiley \& Sons, 2012.

\bibitem{Cioffi91MC}
J.~M. Cioffi, ``A multicarrier primer,'' \emph{ANSI T1E1. Committee
  Contribution}, 1991. 

\bibitem{mukherjee12detecting}
A.~Mukherjee and A.~L. Swindlehurst, ``Detecting passive eavesdroppers in the
  {MIMO} wiretap channel,'' in \emph{Proc. Int. Conf. Acoustics, Speech and
  Signal Processing (ICASSP)}, Tokyo, Japan, Mar. 2012.

\bibitem{Floudas1995MINLP}
C.~A. Floudas, \emph{Nonlinear and {M}ixed {I}nteger {O}ptimization:
  {F}undamentals and {A}pplications}.\hskip 1em plus 0.5em minus 0.4em\relax
  Oxford University Press, 1995.

\bibitem{Bertsekas82Lagrange}
D.~P. Bertsekas, \emph{Constrained {O}ptimization and {L}agrange {M}ultiplier
  {M}ethods}.\hskip 1em plus 0.5em minus 0.4em\relax Academic Press, 1982.

\bibitem{Ye2011interior}
Y.~Ye, \emph{Interior {P}oint {A}lgorithms: {T}heory and {A}nalysis}.\hskip 1em
  plus 0.5em minus 0.4em\relax John Wiley \& Sons, 1997.

\bibitem{cvx}
M.~Grant and S.~Boyd, ``{CVX}: Matlab software for disciplined convex
  programming, version 2.1,'' [Online] Available: \url{http://cvxr.com/cvx},
  Dec. 2016.

\bibitem{polik10interior}
I.~P{\'o}lik and T.~Terlaky, ``Interior point methods for nonlinear
  optimization,'' in \emph{Nonlinear {O}ptimization}.\hskip 1em plus 0.5em
  minus 0.4em\relax Springer Berlin Heidelberg, 2010, pp. 215--276.

\bibitem{Zhao13TWC:BchOpt}
J.~Zhao, T.~Q. Quek, and Z.~Lei, ``Coordinated multipoint transmission with
  limited backhaul data transfer,'' \emph{{IEEE} Trans. Wireless Commun.},
  vol.~12, no.~6, pp. 2762--2775, Jun. 2013.

\bibitem{Zhuang14TSP:AsyCoop}
F.~Zhuang and V.~K. Lau, ``Backhaul limited asymmetric cooperation for {MIMO}
  cellular networks via semidefinite relaxation,'' \emph{{IEEE} Trans. Signal
  Process.}, vol.~62, no.~3, pp. 684--693, Feb. 2014.

\bibitem{ScutariTSP15:SCP}
S.~Sardellitti, G.~Scutari, and S.~Barbarossa, ``Joint optimization of radio
  and computational resources for multicell mobile-edge computing,''
  \emph{{IEEE} Trans. Signal Inf. Process. Netw.}, vol.~1, no.~2, pp. 89--103,
  Jun. 2015.

\bibitem{3GPP:TR36814}
{3GPP TR 36.814}, ``Further advancements for {E-UTRA} physical layer aspects
  ({R}elease 9),'' Mar. 2010.

\bibitem{Horn2012Matrix}
R.~A. Horn and C.~R. Johnson, \emph{Matrix {A}nalysis}.\hskip 1em plus 0.5em
  minus 0.4em\relax Cambridge University Press, 2012.

\bibitem{Luo04quadratic}
Z.-Q. Luo, J.~F. Sturm, and S.~Zhang, ``Multivariate nonnegative quadratic
  mappings,'' \emph{{SIAM} J. Optimization}, vol.~14, no.~4, pp. 1140--1162,
  Jul. 2004.

\bibitem{Bertsekas99NLP}
D.~P. Bertsekas, \emph{Nonlinear {P}rogramming}, 2nd~ed.\hskip 1em plus 0.5em
  minus 0.4em\relax Athena Scientific, 1999.

\bibitem{Aubin76:DualGap}
J.~P. Aubin and I.~Ekeland, ``Estimates of the duality gap in nonconvex
  optimization,'' \emph{Mathematics of Operations Research}, vol.~1, no.~3, pp.
  225--245, Aug. 1976.

\bibitem{Bertsekas83:UC}
D.~Bertsekas, G.~Lauer, N.~Sandell, and T.~A. Posbergh, ``Optimal short-term
  scheduling of large-scale power systems,'' \emph{{IEEE} Trans. Autom.
  Control}, vol.~28, no.~1, pp. 1--11, Jan. 1983.

\bibitem{Giannakis13:SG}
S.-J. Kim and G.~Giannakis, ``Scalable and robust demand response with
  mixed-integer constraints,'' \emph{{IEEE} Trans. Smart Grid}, vol.~4, no.~4,
  pp. 2089--2099, Apr. 2013.

\end{thebibliography}

\end{document}